\documentclass[11pt,prd,reprint,showkeys, amsmath,amssymb, aps, nofootinbib, floatfix]{revtex4-2}
\DeclareUnicodeCharacter{2212}{\textendash}
\usepackage{graphicx,subfigure}
\usepackage[export]{adjustbox}
\usepackage{dcolumn}
\usepackage{bm}
\usepackage{physics}
\usepackage{xcolor}
\usepackage{slashed,cancel}
\usepackage{multirow}
\usepackage{capt-of}
\usepackage{comment}
\usepackage{mathrsfs}
\usepackage{enumitem}
\usepackage{siunitx}
\usepackage[colorlinks=true,linkcolor=blue,urlcolor=blue,citecolor=blue]{hyperref}

\allowdisplaybreaks
\begin{document}
\preprint{APS/123-QED}

\title{Constraint on anomalous $W^-W^+\gamma$ coupling in $\ell^\pm \gamma\slashed{E}$ final state at NLO QCD}

\author{Amir Subba}
\email{amirsubba@ustc.edu.cn} 
\author{Yu Shi}%
\email{yu\_shi@ustc.edu.cn}
\affiliation{Wilczek Quantum Center, Shanghai Institute for Advanced Studies, Shanghai 201315, China}
\affiliation{University of Science and Technology of China, Hefei 230026, China }

\author{Athul P}
\email{ap22rs013@iiserkol.ac.in} %
\author{Ritesh K Singh}
\email{ritesh.singh@iiserkol.ac.in} %
\affiliation{Department of Physical Sciences, Indian Institute of Science Education and Research Kolkata, Mohanpur, West Bengal 741246, India}

\date{\today}

\begin{abstract}
We study the anomalous $W^-W^+\gamma$ triple gauge couplings in the $pp\to \ell^\pm\gamma\slashed{E}$ channel at $\sqrt{s}=13$~TeV with next-to-leading order QCD corrections. Eight polarization asymmetries of the reconstructed $W$ boson, measured in four disjoint regions of $\Delta\eta_{\ell\gamma}$ and combined with the $m_{W\gamma} \ge 1$ TeV distribution, are used to constrain the two CP-even and two CP-odd couplings. The bounds improve only marginally with increasing luminosity, far below the statistical expectation, while they degrade substantially when the systematic uncertainties are relaxed, showing that the sensitivity is systematically limited. A simultaneous four-parameter MCMC fit is performed to obtain marginalized limits. Relative to the CMS $WW+WZ$ combination, our bounds are significantly tighter for the CP-odd coupling $\widetilde{\lambda}_\gamma$ and competitive for $\lambda_\gamma$. The CP-odd results are also compared with the electron EDM limit, which constrains both couplings by roughly $2.5$ orders of magnitude beyond the collider reach.
\end{abstract}

\maketitle

\section{Introduction}
\label{sec:intro}

The associated production of a $W$ boson and a hard photon at a hadron collider, $pp \to W^\pm \gamma + X$ is one of the most theoretically rich and phenomenologically important diboson processes accessible at the LHC. At the fundamental level, it directly probes one of the charged triple gauge couplings (TGCs) $W^-W^+\gamma$ which is fixed by $SU(2)_L \times U(1)_Y$ gauge structure of the Standard Model (SM). The $W^-W^+\gamma$ vertex encodes the electromagnetic multipole structure of the $W$ boson. At the leading order (LO), the gauge invariance fixes the magnetic dipole moment of $W$ boson $\mu_W = e/m_W $ and electric quadrupole moment $Q_W = -e/m_W^2$~\cite{Salzman1966}. Further, the  process is also of special interest due to the existence of radiation-amplitude zero (RAZ)~\cite{Brown:1979ux,Mikaelian:1979nr,Goebel:1980es,Brown:1982xx,Brodsky:1982sh} at the LO amplitude of the parton level subprocesses. Crucially, the position of this zero is independent of the photon energy and depends only on the fractional electric charges of the quarks~\cite{Samuel:1997zs}. 

Thus the $W\gamma$ process becomes highly sensitive to the deviation in the $W^-W^+\gamma$ couplings. Nevertheless, the higher order quantum corrections could also bring change in the structure of the RAZ and polarizations of the $W$ boson. Hence, the analysis performed at the LO may not break the degeneracy between the anomalous nature and quantum corrections.

The $W\gamma$ process also allows us to exploit additional sensitive observables in the form of eight independent polarizations of the $W$ boson, which can be obtained from the angular distribution of the final decayed charged fermion in the rest frame of the $W$ boson~\cite{Aguilar-Saavedra:2015yza,Rahaman:2016pqj} which serve as sensitive probes of new physics~\cite{Subba:2022czw,Subba:2023jia,Subba:2023rpm,Subba:2024ojy,Subba:2025hkq,Subba:2025hxn,Subba:2025pos,Eboli:2026mbq,Rahaman:2016pqj,Rahaman:2017qql,Rahaman:2019lab,Rahaman:2019mnz,Rahaman:2021fcz,Rahaman:2023pte}. Next-to-leading (NLO) QCD corrections to the $W$ polarizations along with spin correlations in presence of anomalous TGC have been reported~\cite{Pelliccioli:2026ltl}. 

The $W\gamma$ process is one of the most precisely predicted and theoretically well-understood diboson processes. The NLO QCD corrections to $W\gamma$ production with the on-shell $W$ bosons were evaluated in Refs.~\cite{Smith:1989xz,PhysRevD.47.940} and later generalized in Ref.~\cite{Dixon:1998py} to account for leptonic decays using the narrow-width approximation and in the presence of anomalous $W^-W^+\gamma$ couplings ~\cite{Baur:1993ir}. The effect of one-loop logarithmic electroweak (EW) radiative corrections on $W\gamma$ production processes at the LHC has been studied ~\cite{Accomando:2001fn}. The calculation of the full electroweak corrections to $W\gamma$ production, including the decays of the W boson in the pole approximation, was presented in Ref.~\cite{Accomando:2005ra}. The complete NLO QCD+EW calculation taking into account all off-shell effects of $W$ with the finite width of the $W$ boson is done in Ref.~\cite{Denner:2014bna}. The $W\gamma$ process has been computed in next-to-NLO QCD in Refs.~\cite{Grazzini:2015nwa, Grazzini:2016hai,Campbell:2021mlr}.

On the experimental side, the first verification of the RAZ along with constraints on the anomalous $W^-W^+\gamma$ couplings has been provided by CDF~\cite{CDF:2004tuf} and D0~\cite{D0:2005djn,D0:2008abl} collaborations at the Fermi\-lab Tevatron in $p\bar{p}$ collisions at $\sqrt{s} = 1.96$~TeV, using leptonic decay modes of the $W$ boson. At the LHC, the $W\gamma$ process has been studied at $\sqrt{s} = 7$~TeV by both CMS~\cite{CMS:2011myh} and ATLAS~\cite{ATLAS:2012bpb,ATLAS:2012upi,ATLAS:2013way,CMS:2013ryd} and more recently at $\sqrt{s} = 13$~TeV~\cite{CMS:2021foa,CMS:2021cxr}.
From the theoretical perspective, the $W^-W^+\gamma$ vertex has been extensively investigated in the literature across a variety of collider environments, including the LHC~\cite{Baur:1988qt,Papavassiliou:1999iv,Etesami:2016rwu,He:1993hx,Bhatia:2018ndx,PhysRevD.51.90,Tizchang:2020tqs,Yang:2012vv,Bobeth:2015zqa,Etesami:2016rwu,Dawson:2013owa,Eboli:2024wgy}, lepton colliders~\cite{Hagiwara:1986vm,PhysRevD.107.073004,Subba:2023rpm,CHOUDHURY1999193,Ari:2015tca,Jahedi:2024wnw}, as well as various other collider setups~\cite{Baur:1993ix,Baur:1989gh,Baur:1989gk,Biswal:2014oaa,Kumar:2015lna,Li:2017kfk,Sahin:2017rfr,rizzo1999polarizationasymmetriesgammae,Biswal:2014oaa,Dawson:1996ge,Gutierrez-Rodriguez:2019hek,Koksal:2019oqt}.

Though, there exists a large body of literature that has focused on probing anomalous TGC in $W\gamma$ process, the work that takes into account all eight polarizations of $W$ boson combined with RAZ phenomena in presence of next-to-leading corrections in QCD is still lacking.  The current work tries to fill that phenomenological gap by studying $W$ boson polarization in disjoint region of RAZ along with reconstructed $m_{W\gamma}$ at NLO QCD in presence of anomalous $W^-W^+\gamma$ coupling. We performed our analysis via $pp\to \ell^\pm \gamma \slashed{E}$ process at $\sqrt{s}=13$ TeV. The $W$ boson is reconstructed using on-shell constraints of $W$ boson and the missing transverse momenta.

In the current work, we use a most general Lorentz invariant and $U(1)_{em}$ invariant $W^-W^+\gamma$ vertex parametrized through an effective Lagrangian~\cite{Hagiwara:1986vm}
\begin{align}
	\label{eq:Lageff}
	\mathscr{L}_{WW\gamma} &= -ig_W\sin\theta_W\Big[g_1^\gamma(W^+_{\mu \nu}W^{-\mu} - W^{+\mu}W^-_{\mu \nu})A^\nu \nonumber\\&+  k_\gamma W^+_\mu W^-_\nu A^{\mu \nu} +\frac{\lambda_\gamma}{m_W^2}W_\mu^{\nu+}W_\nu^{-\rho}A_{\rho}^{\mu} \nonumber\\&+ig_4^\gamma W_\mu^+W_\nu^-(\partial^\mu A^\nu+\partial^\nu A^\mu) \nonumber\\&- ig_5^\gamma \epsilon^{\mu \nu \rho \sigma}(W_\mu^+ \partial_\rho W_\nu^- - \partial_\rho W_\mu^+W_\nu^-)A_\sigma  
	\nonumber\\&+\tilde{k}_\gamma W_\mu^+W_\nu^-\tilde{A}^{\mu \nu} + \frac{\tilde{\lambda}_\gamma}{m_W^2}W_\mu^{\nu+}W_\nu^{-\rho}\tilde{A}_\rho^{\mu}\Big],
\end{align}
where the field-strength tensors are defined as $W^\pm_{\mu\nu}=\partial_\mu W^\pm_\nu-\partial_\nu W^\pm_\mu$ and $A_{\mu\nu}=\partial_\mu A_\nu-\partial_\nu A_\mu$, and the dual tensor is $\widetilde{A}^{\mu\nu}=\tfrac{1}{2}\epsilon^{\mu\nu\rho\sigma}A_{\rho\sigma}$ with the convention $\epsilon^{0123}=+1$. 

Within the SM at tree level, the parameter $g_1^\gamma = 1$ determines the charge of the $W$ boson and the deviation in $g_1^\gamma$ is null due to the electromagnetic gauge invariance. The parameter $\kappa_\gamma = 1$  defines the magnetic moment of the $W$ boson. All other couplings are zero within the SM tree level. The couplings $g_1^\gamma$, $\kappa_\gamma$ and $\lambda_\gamma$ are even under both parity (P) and charge conjugation (C), while $g_4^\gamma$ is C-odd and P-even. The $\widetilde{\kappa}_\gamma$ and $\widetilde{\lambda}_\gamma$ are C-even and P-odd making them odd under combined CP transformation. Further, imposing $SU(2)\times U(1)$ gauge invariance, the two couplings $g_4^\gamma$ and $g_5^\gamma$ vanish.  The remaining anomalous couplings are directly linked to the electromagnetic multipole moments of the $W$ boson: $\delta\kappa_\gamma\equiv \kappa_\gamma-1$ parameterizes a deviation in the magnetic dipole moment while $\lambda_\gamma$ modifies the electric quadrupole moment. The CP-violating couplings $\widetilde\kappa_\gamma$ and $\widetilde\lambda_\gamma$ are associated with the electric dipole ($d_W$) and magnetic quadrupole ($\widetilde{Q}_W$) moments of the $W$, respectively~\cite{Hagiwara:1986vm}
\begin{align}
    d_W = \frac{e}{2\,m_W}(\widetilde{\kappa}_\gamma + \widetilde{\lambda}_\gamma),\quad \widetilde{Q}_W = -\frac{e}{m_W^2}(\widetilde{\kappa}_\gamma - \widetilde{\lambda}_\gamma).
\end{align}
The moments $d_W$ and $\widetilde{Q}_W$ are expected to be extremely small within the SM, for e.g, $d_W$ remains zero to two loops~\cite{Booth:1993af,Shabalin:1978rs} and receive first non-zero contribution at three loops~\cite{Pospelov:1991zt}. The $\widetilde{Q}_W$ of the $W$ boson receives a tiny contribution at the two–loop level in the SM~\cite{Khriplovich:1993js}. Using the neutron dipole moment $(d_n)$, the bounds on the CP-violating couplings $\widetilde{\kappa}_\gamma$ and $\widetilde{\lambda}_\gamma$ are obtained in Refs.~\cite{Montano:2005kt}. We revisit the sensitivity of $pp\to \ell\gamma\slashed{E}$ process to these CP-violating couplings with polarization asymmetries of $W$ boson and $m_{W\gamma}$ distribution.  

The paper is organized as follows. In sec.~\ref{sec:pol}, we discuss the polarizations of the $W$ boson along with its reconstruction using the asymmetries in angular functions of final charged fermions. In sec.~\ref{sec:collider}, we discuss the simulation methodology along with event selection and neutrino reconstruction method. In sec.~\ref{sec:results}, we investigate the impact of anomalous couplings on various kinematic distributions and derive the corresponding constraints on these couplings by exploiting the polarizations of reconstructed $W$ boson for different integrated luminosities and systematic uncertainties. In sec.~\ref{sec:mcmc}, we perform a  global analysis of aTGC using Markov-Chain-Monte-Carlo method.  We discuss the electron electric dipole moment in sec.~\ref{sec:edm}. Finally, we conclude in sec.~\ref{sec:con}.

\section{Polarization of $W$ boson}
\label{sec:pol}
The $W$ boson is a massive spin-1 particle and its spin state is characterized by a $3\times 3$ Hermitian density matrix $\rho_{\lambda\lambda'}$, where $\lambda,\lambda'\in\{+1,\,0,\,-1\}$ are $W$ helicity states.  Being Hermitian with unit trace, the density matrix has eight independent real degrees of freedom.  Following the multipole decomposition of Refs.~\cite{Boudjema:2009fz,Aguilar-Saavedra:2015yza,Rahaman:2016pqj}, these eight degrees of freedom can be organized into two sets viz., three \emph{vector polarizations} (corresponding to the expectation values $\langle J_x\rangle$, $\langle J_y\rangle$, $\langle J_z\rangle$ of the angular-momentum operators in the $W$ rest frame), and five independent components of the symmetric, traceless rank-2 \emph{tensor polarization} $T_{ij}$, where $i,j\in\{x,y,z\}$.  

The polarization of the $W$ boson can be experimentally accessed through the angular distribution of its decay products. For the leptonic decay $W^- \to \ell^- \bar\nu_\ell$ (and its charge conjugate), the charged lepton serves as the spin analyzer owing to the maximal parity violation of the $V-A$ weak interaction. In the rest frame of the $W$ boson, the full angular distribution of the decay lepton reads~\cite{Boudjema:2009fz}
\begin{align}
  \label{eqn:diffn}
  \frac{1}{\sigma}\frac{d\sigma}{d\Omega_\ell} &=
    \frac{3}{8\pi}\Bigg[
      \left(\frac{2}{3} - (1-3\delta)\,\frac{T_{zz}}{\sqrt{6}}\right)
      + \alpha\,P_z\cos\theta_\ell \nonumber\\
    &+ \sqrt{\frac{3}{2}}\,(1-3\delta)\,T_{zz}\cos^2\!\theta_\ell \nonumber\\
    &+ \left(\alpha\,P_x + 2\sqrt{\frac{2}{3}}\,(1-3\delta)\,T_{xz}\cos\theta_\ell\right)
       \sin\theta_\ell\cos\phi_\ell \nonumber\\
    &+ \left(\alpha\,P_y + 2\sqrt{\frac{2}{3}}\,(1-3\delta)\,T_{yz}\cos\theta_\ell\right)
       \sin\theta_\ell\sin\phi_\ell \nonumber\\
    &+ (1-3\delta)\,\frac{T_{xx}-T_{yy}}{\sqrt{6}}\,\sin^2\!\theta_\ell\cos(2\phi_\ell)
       \nonumber\\
    &+ \sqrt{\frac{2}{3}}\,(1-3\delta)\,T_{xy}\sin^2\!\theta_\ell\sin(2\phi_\ell)
    \Bigg],
\end{align}
where $\theta_\ell$ and $\phi_\ell$ are the polar and azimuthal angles of the charged lepton in the $W$ rest frame.  The parameters $\alpha$ and $\delta$ encode the underlying $V$-$A$ coupling structure and are given by~\cite{Boudjema:2009fz}
\begin{align}
  \label{eqn:alphadelta}
  \alpha &= \frac{2(C_R^2 - C_L^2)\sqrt{1 + (x_1^2-x_2^2)^2 - 2(x_1^2+x_2^2)}}
               {12 C_L C_R x_1 x_2 + G(C_R^2+C_L^2)}, \nonumber\\
  \delta &= \frac{4 C_L C_R x_1 x_2 + (C_R^2+C_L^2)(G-2)}
               {12 C_L C_R x_1 x_2 + G(C_R^2+C_L^2)},
\end{align}
where $G = 2 - (x_1^2-x_2^2)^2 + (x_1^2+x_2^2)$, $x_i = m_i / m_W$ are the scaled masses of the decay fermions, and $C_L$, $C_R$ are the left- and right-chiral couplings of the $W$ to the final state lepton.  In the massless-fermion limit relevant for $W\to \ell\nu$, $x_1, x_2 \to 0$, one finds $\delta \to 0$ and $\alpha \to (C_R^2 - C_L^2)/(C_R^2 + C_L^2)$.  Within the SM, the $W$ boson couples exclusively to left-handed fermions ($C_R = 0$), giving $\alpha = -1$.  The parameter $\alpha = -1$ is therefore the \emph{spin-analyzing power} of the charged lepton, which takes its maximal absolute value owing to the purely $V-A$ structure; this makes leptonic $W$ decays the ideal analyzer for the full spin state.

Each of the eight polarizations are directly accessible from a well-defined angular asymmetry of the lepton distribution, constructed by integrating the differential cross section over the appropriate phase-space regions. For example, the $P_z$ and $T_{xy}$ can be obtained in the form of appropriate asymmetries as~\cite{Rahaman:2016pqj,Rahaman:2021fcz}
\begin{align}
&\mathcal{A}_z = \left(\int_{\theta_\ell=0}^{\pi/2}\int_{\phi_\ell=0}^{2\pi} - \int_{\theta_{\ell}=\pi/2}^\pi\int_{\phi_\ell = 0}^{2\pi}\right)d\Omega_\ell\left(\frac{1}{\sigma}\frac{d\sigma}{d\Omega_\ell}\right),\nonumber\\
&= \frac{1}{2}\alpha\,P_z,\nonumber\\
    &\mathcal{A}_{xy} = \left(\int_{\theta_\ell=0}^\pi\int_{\phi_\ell=0}^{\pi/2}-\int_{\theta_{\ell}=0}^{\pi/2}\int_{\phi_\ell=\pi/2}^\pi\nonumber\right.\\&+\left. \int_{\theta_{\ell}=\pi/2}^\pi\int_{\phi_\ell = \pi/2}^{3\pi/2}-\int_{\theta_\ell=\pi/2}^\pi\int_{\phi_\ell=-\pi/2}^{\pi/2}\right)d\Omega_\ell \left(\frac{1}{\sigma}\frac{d\sigma}{d\Omega_\ell}\right),\nonumber\\
    &= \frac{2}{\pi}\sqrt{\frac{2}{3}}\,(1-3\,\delta)\,T_{xy}.
\end{align}

The eight polarizations transform differently under parity and charge conjugation. The three vector polarizations are odd under parity while the remaining five tensorial are even. Under combined CP transformation, $P_y$, $T_{xy}$ and $T_{yz}$ are odd while rest are even.

In the case of $pp\to \ell\nu\gamma$ process, the QCD radiative corrections enter exclusively at the level of the hard production subprocess $q\bar{q}'\to W\gamma$ and its higher-order partonic channels, while the leptonic decay $W\to\ell\nu$ proceeds through purely electroweak dynamics and is unaffected by QCD at any order.  As a consequence, the analytic form of Eq.~\eqref{eqn:diffn} is preserved to all orders in $\alpha_s$.

The CP-odd asymmetries which are sensitive to the imaginary parts of the off-diagonal density matrix elements first arise at $\mathcal{O}(\alpha_s)$ in the SM from absorptive contributions to the production amplitude~\cite{Mirkes:1992hu}.

In our analysis, the $W$ four-momentum is reconstructed from the charged lepton and the missing transverse energy, using the $W$ mass constraint $m_W^2 = (p_\ell + p_\nu)^2$ to resolve the two-fold ambiguity in the longitudinal neutrino momentum. The reconstruction of the neutrino is discussed in the next section.
\section{Collider analysis and Neutrino reconstruction}
\label{sec:collider}
The analysis is performed for proton-proton collisions at a center-of-mass energy of $\sqrt{s}=13$~TeV. The effective Lagrangian defined in Eq.~\eqref{eq:Lageff} is implemented in \textsc{Feynrules}~\cite{Alloul:2013bka} to generate a \textsc{Universal FeynRules Output} (UFO~\cite{Degrande:2011ua,Darme:2023jdn}) model file. Next-to-leading order (NLO) QCD corrections are incorporated into the model, with the calculation of counter-terms and rational terms carried out using \textsc{NLOCT}~\cite{Degrande:2014vpa}.

Events are generated with  \textsc{MadGraph5$\_$aMC$@$NLO} \cite{Alwall:2014hca} at NLO accuracy in QCD. We use \textsc{NNPDF23} set for the parton distribution function (PDFs) with $\alpha_s(m_Z)=0.118$. The generated parton-level events are passed to \textsc{Pythia}~\cite{Bierlich:2022pfr} for showering and hadronization. Detector response, including geometric acceptance, reconstruction efficiencies, and resolution effects, is subsequently modeled using \textsc{Delphes}~\cite{deFavereau:2013fsa} with the default CMS detector card.

Jets are clustered with \textsc{anti-$k_T$}~\cite{Cacciari:2008gp} clustering algorithm with jet radius $R=0.4$. Final state charged leptons and photon undergo an isolation criteria defined as~\cite{deFavereau:2013fsa}
\begin{align}
    I(P) = \frac{\sum_{i\neq P}^{\Delta R < R,\,p_T(i) > p_T^{\mathrm{min}}}p_T(i)}{p_T(P)},
\end{align}
where $P \in \{\ell,\gamma\}$, and the numerator is the sum of transverse momenta above $p_T^{\mathrm{min}}$ of all particles that lie within a cone of radius $R$ around the lepton or photon. For both charged leptons and photon, we choose $R = 0.5$, and $p_T^{\mathrm{min}} = 0.5$ GeV. For the isolation condition,  we choose $I(\ell)^{\mathrm{max}} = 0.12\,(0.25)$ for electron (muon) and $I(\gamma)^{\mathrm{max}} = 0.12$.
\begin{figure*}[!htb]
    \centering
    \includegraphics[width=0.49\textwidth]{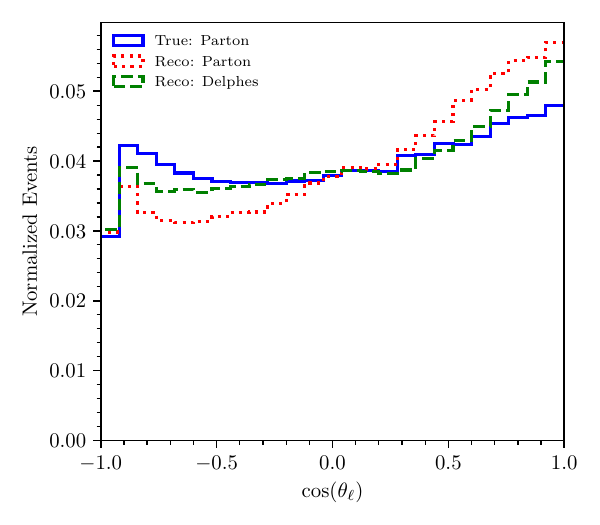}
    \includegraphics[width=0.49\textwidth]{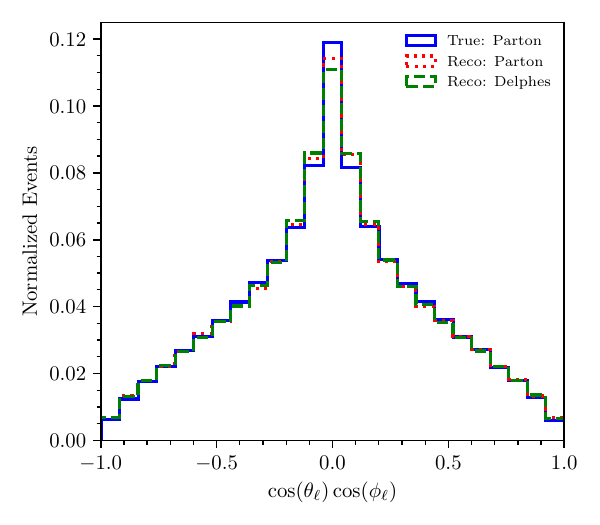}
   \caption{\label{fig:angfunc}
Normalized distributions of $\cos\theta$ (left) and $\cos\theta\cos\phi$ (right) of charged lepton comparing parton-truth, parton-level reconstruction, and \textsc{Delphes} reconstruction. The $\cos\theta$ distribution shows a systematic shift at reconstruction levels, while $\cos\theta\cos\phi$ exhibits near-perfect agreement.
}
\end{figure*}

We implement the event selection criteria of CMS in Ref.~\cite{CMS:2021cxr} as follows. Events with isolated leptons and photons having $p_T^\ell \ge 25.0$ GeV, $|\eta_\ell| < 2.5$, $p_T^\gamma > 30.0$~GeV, $|\eta_\gamma| < 2.5$ and $\Delta R_{\gamma\ell} \equiv \sqrt{(\Delta\eta_{\gamma\ell})^2+(\Delta\phi_{\gamma\ell})^2}$ $> 0.7$ are selected. Hadronic jets are counted if $p_T^j \ge 30$~GeV, $|\eta_j| < 2.5$ and separation from both lepton and photon of $\Delta R > 0.4$. In order to reduce background contribution from processes like $Z \to \ell^-\ell^+$, we set $\slashed{E}_T > 30$ GeV and events with $m_{\ell\gamma}$, the invariant mass of lepton-photon pair, close to $m_Z$ i.e $70 < m_{\ell \gamma} < 100$ GeV are rejected.

We generate the matrix-element-level events with MG5 using the following kinematic cuts
\begin{align}
    p_T^\ell \ge 25.0~\mathrm{GeV},\quad &p_T^\gamma \ge 30.0~\mathrm{GeV},\quad |\eta_\ell| \le 2.5, \nonumber\\ &|\eta_\gamma| \le 2.5,
\end{align}
where $p_T^{\ell/\gamma}$ and $\eta_{\ell/\gamma}$ are the transverse momenta and pseudorapidities of the lepton and photon, respectively. 

We consider three major backgrounds viz. $Z\gamma$, $t\bar{t}\gamma$ and $Vj,~V\in\{W,Z\}$ process at the leading order. The detector efficiency along with the expected number of events at an integrated luminosity $\mathcal{L}=140$ fb$^{-1}$ are listed in Table~\ref{tab:sigbkg}. The background rates are normalized to NLO QCD by multiplying by the relevant $K$-factor, with $K_{Z\gamma} = 1.46$ and $K_{t\bar{t}\gamma}=1.45$~\cite{Alwall:2014hca}.  The $Vj$ background process is removed from further analysis due to its negligible selection rate.
\begin{table}[!htb]
    \centering
    \renewcommand{\arraystretch}{1.4}
    \caption{Cut-flow chart of survival fraction of events after different kinematic cuts for signal and various background processes and expected number of events at an integrated luminosity of 140 fb$^{-1}$. The background rates are normalized to the NLO QCD by multiplying with NLO QCD $K$-factor.}
    \label{tab:sigbkg}
    \begin{tabular*}{0.48\textwidth}{@{\extracolsep{\fill}}lcc@{}}\hline
         Process & $\epsilon_D = N_{\mathrm{Cuts}}/N_{\mathrm{MC}}$ &$N(\mathcal{L}=140~\mathrm{fb}^{-1})$\\ \hline
          $\ell \gamma \slashed{E}$&$0.281$ & $2.0\times 10^5$ \\
          $Z\gamma$&$0.003$&$1.5\times 10^4$\\
          $t\bar{t}\gamma$ & $0.004 $&$974$ \\
          $Vj$&$\sim 0$& \\

        \hline
    \end{tabular*}
\end{table}
\\
\noindent
\textbf{Neutrino reconstruction:} We consider the leptonic decay of a $W$ boson:
\begin{align}
W \to \ell \nu,
\end{align}
where $\ell = e,\mu$ and $\nu$ is the corresponding neutrino. The $W$ boson is assumed to be on-shell, the lepton and neutrino are treated as massless, and the neutrino is the only invisible particle in the event, carrying missing transverse energy.
\begin{figure*}[!htb]
    \centering
    \includegraphics[width=0.49\linewidth]{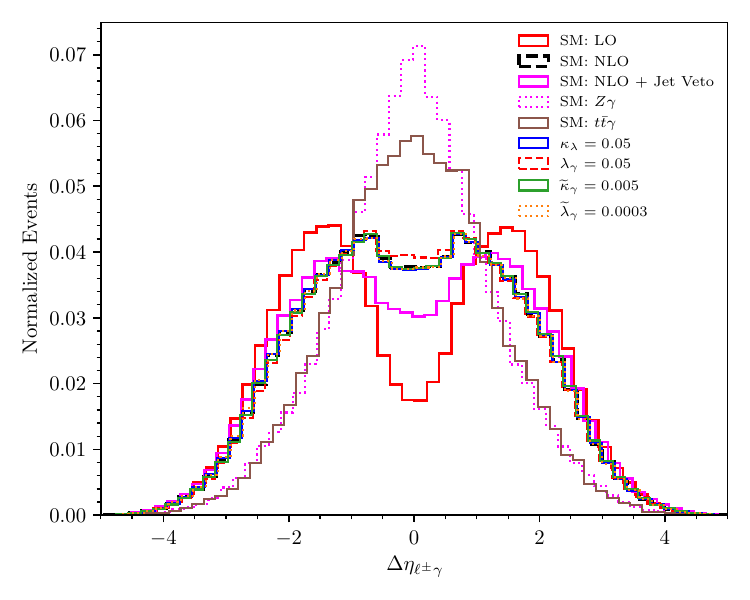}
    \includegraphics[width=0.49\textwidth]{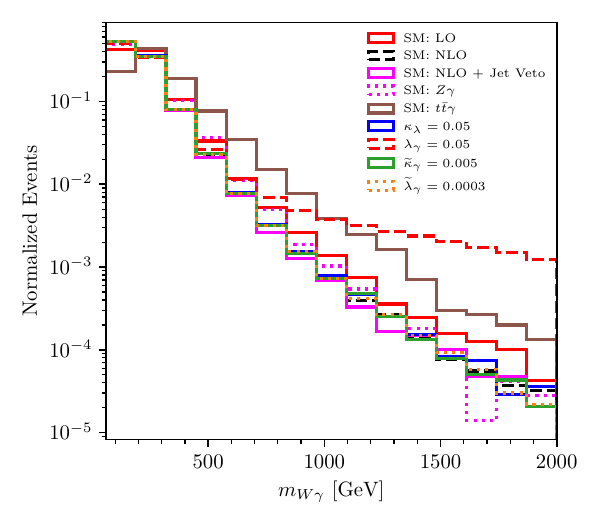}
    \caption{Normalized distributions of pseudorapidity difference $\Delta\eta_{\ell^\pm\gamma}$ and reconstructed invariant mass of $W$ and photon system $(m_{W\gamma})$ for SM predictions and the anomalous coupling benchmarks.}
    \label{fig:kin}
\end{figure*}

The on-shell condition $p_W^2 = m_W^2$ yields two solutions for the neutrino longitudinal momentum:
\begin{align}
p_{z,\nu}^{\pm} = \frac{ \beta \, p_{z\ell} \pm E_\ell \sqrt{ \beta^2 - p_{T\nu}^2 \, p_{T\ell}^2 } }{ p_{T\ell}^2 },
\end{align}
with $\beta = \frac{m_W^2}{2} + p_{T\ell}\cdot p_{T\nu}$. A real solution requires $\beta \ge p_{T\nu}p_{T\ell}$. When the discriminant is negative we sample $m_W$ from a normal distribution $\mathcal{N}(80.4, 2.0)$ for up to $1000$ iterations; if no real solution is found, the discriminant is set to zero. Of the two solutions, we select the one that minimizes the reconstructed $m_W$ error.

We compare the angular functions $\cos\theta$ and $\cos\theta\cos\phi$, which are sensitive to the vector ($P_z$) and tensor ($T_{xz}$) polarizations of the $W$ boson, respectively. The corresponding distributions are shown in Fig.~\ref{fig:angfunc}. For $\cos\theta$ (left panel), both reconstruction levels exhibit a moderate deviation from the parton-truth: the event yield is suppressed at negative $\cos\theta$ and enhanced at positive $\cos\theta$ relative to the truth. This asymmetry distortion suggests a systematic shift in the reconstructed lepton angular distribution, which would propagate into the extracted polarization parameters. In contrast, the $\cos\theta\cos\phi$ distribution (right panel) shows excellent agreement across all three levels, with the parton and \textsc{Delphes} reconstruction curves nearly overlapping the truth. This indicates that the reconstruction preserves the azimuthal correlation well, and the associated tensor polarization $T_{xz}$ remains robust against detector effects and kinematic ambiguities.
\section{Limits on anomalous couplings}
\label{sec:results}
In order to highlight the impact of anomalous couplings on the kinematic distributions, we consider a benchmark point of $0.05$ for $\delta\kappa_\gamma$ and $\lambda_\gamma$, $\widetilde{\kappa}_\gamma = 0.005$ and $\widetilde{\lambda}_\gamma=  0.0003$. We consider two kinematic distributions (see Fig.~\ref{fig:kin}) viz., the pseudorapidity difference between the lepton and the photon ($\Delta\eta_{\ell^\pm\gamma}$) and the reconstructed invariant mass of $W$ and photon system $m_{W\gamma}$.

The distribution of $\Delta\eta_{\ell^\pm\gamma}$ exhibits a pronounced dip at zero for the leading-order (LO) SM events (solid red curve), consistent with the RAZ phenomenon. This dip is lifted by NLO QCD corrections (dashed black curve), which introduce additional jet activity that fills the zero region. When a jet veto is applied at NLO (solid magenta curve), the dip is partially restored, indicating that hard jet radiation is primarily responsible for filling the RAZ. The dominant backgrounds, $Z\gamma$ (dotted magenta curve) and $t\bar{t}\gamma$ (solid brown curve), do not exhibit a RAZ, instead showing a peak around zero. Notably, the distributions for all the anomalous couplings closely resemble the NLO QCD prediction, with a moderately filled dip. The $m_{W\gamma}$ distribution shows a falling spectrum for all SM predictions, with aTGC enhancing the high-$m_{W\gamma}$ tail. The $\lambda_\gamma$ shows a pronounced distribution in the tail-region  due to the $(\sqrt{s}/2m_W)^2$ dependence of anomalous amplitudes compared to $\kappa_\gamma$ which contributes at the linear level~\cite{Baur:1988qt}. 
\begin{figure*}[!htb]
    \centering
    \includegraphics[width=0.49\textwidth]{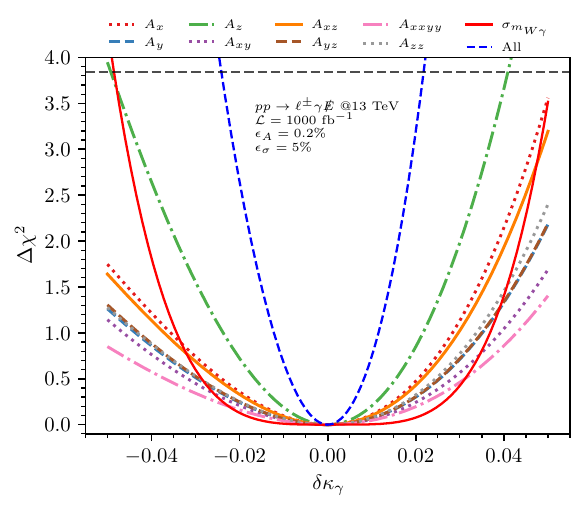}
    \includegraphics[width=0.49\textwidth]{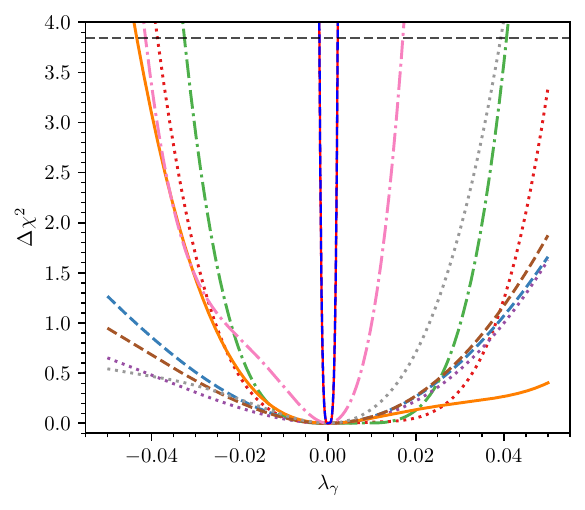}
    \includegraphics[width=0.49\textwidth]{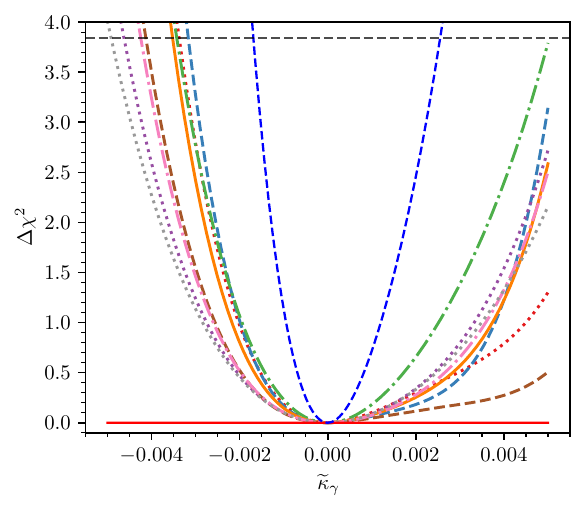}
    \includegraphics[width=0.49\textwidth]{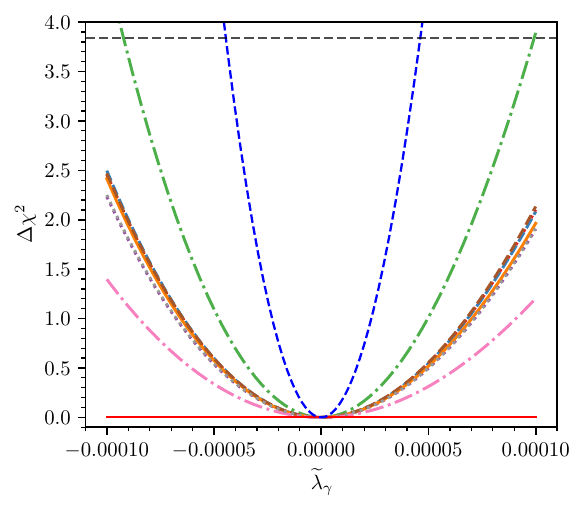}
    \caption{\label{fig:chi_asymm} $\Delta\chi^2$ for set of eight asymmetries and $\sigma_{m_{W\gamma}}$ as a function of one anomalous coupling at a time while others are set to zero. The distributions are obtained with $pp\to \ell^\pm\gamma\slashed{E}$ final states at $\sqrt{s}=13$ TeV, $\mathcal{L}=1000$ fb$^{-1}$ and systematic error ($\epsilon_{\mathcal{A}}, \,\epsilon_\sigma$) of ($0.2\%,\,5\%$).}
\end{figure*}

We further binned the events in four disjoint regions of $\Delta\eta_{\ell\gamma}$ defined as
\begin{align}
    \Delta\eta_{\ell\gamma} \in \Big[\infty,\,-2.0,\,0.0,\,2.0,\,+\infty\Big].
\end{align}
In these four regions, we compute the eight asymmetries of the reconstructed $W$ boson as discussed in the previous section. This set of observables is further used to constrain the aTGC following a chi-squared analysis defined by
\begin{align}
    \label{eq:chi}
    \Delta\chi^2(f) = \sum_{b=1}^4\sum_{i=1}^8\frac{\left(\mathcal{A}_i(f)-\mathcal{A}_i(0)\right)^2}{\delta\mathcal{A}_i^2},
\end{align}
where the indices $i$ and $b$ run over the eight asymmetries and the four regions of $\Delta\eta_{\ell\gamma}$, respectively. Here, $\delta\mathcal{A} = \sqrt{(1-A^2)/(\mathcal{L}\sigma) + \epsilon_A^2}$ is the error in the estimation of asymmetries with $\epsilon_A$ being the systematic error. The Eq.~\eqref{eq:chi} is supplemented by the cross~section for reconstructed $m_{W\gamma} \ge 1$ TeV ($\sigma_{m_{W\gamma}}$), with error given by $\delta\sigma = \sqrt{\sigma/\mathcal{L} + (\epsilon_\sigma\sigma)^2}$. The $\sigma_{m_{W\gamma}}$ is parametrized in presence of anomalous couplings $f$ as
\begin{align}
    \label{eq:xsec}
    \sigma_{m_{W\gamma}}(f_i,f_j) &= \sigma_0 + \sum _i \sigma_i f_i + \sum_{i,j\neq i}\sigma_{ij}f_if_j \nonumber\\&+ \sum_i \sigma_{ii}f_i^2,
\end{align}
where $\sigma_0$ is the SM contribution, $\sigma_i$ is the interference with the SM, $\sigma_{ij}$ represent insertion of two different anomalous couplings and $\sigma_{ii}$ is the quadratic contribution from anomalous amplitude.

We begin by comparing the sensitivity of each asymmetry and $\sigma_{m_{W\gamma}}$ along with their combinations to one aTGC at a time while keeping others to zero at an integrated luminosity $\mathcal{L}=1000$ fb$^{-1}$ with a flat systematic error $\epsilon_{\mathcal{A}}=0.2\%$ and $\epsilon_\sigma=5\%$. The corresponding distributions are shown in Fig.~\ref{fig:chi_asymm}, where the horizontal line at $\Delta\chi^2\sim 3.84$ represents the limits of aTGC with $95\%$ confidence level (CL). For all the aTGCs except $\lambda_\gamma$, we note that the parameter $A_z$ defined as an asymmetry in the $\cos\theta$ distribution of the lepton provides the dominant contribution to the $\chi^2$. While for $\lambda_\gamma$, the $\sigma_{m_{W\gamma}}$ provides the dominant contribution to $\chi^2$ followed by the asymmetry in the $\cos(2\phi)$ distribution of the charged lepton. 

\begin{figure*}[!htb]
    \centering
    \includegraphics[width=0.32\textwidth]{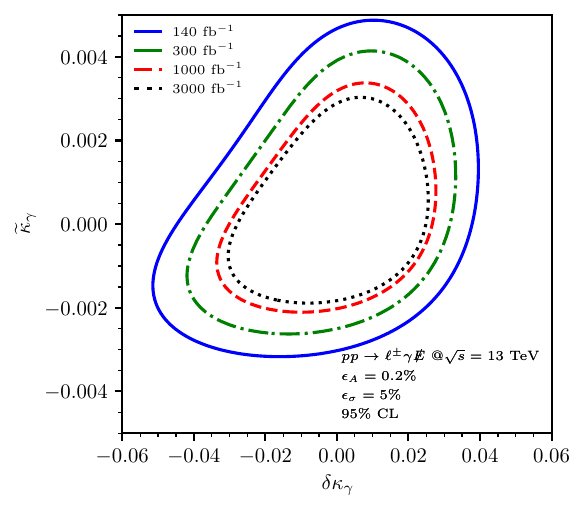}
    \includegraphics[width=0.32\textwidth]{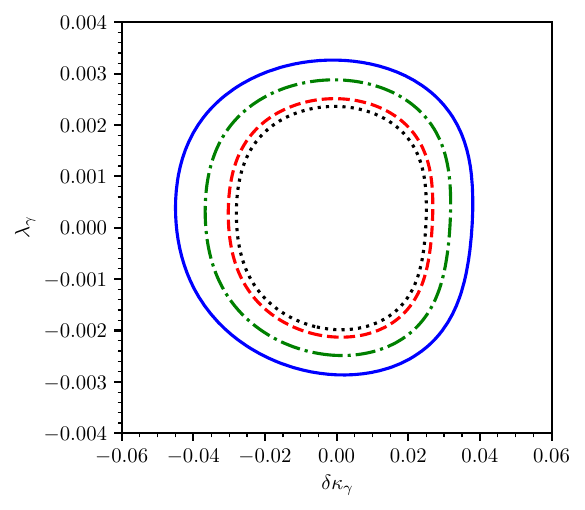}
    \includegraphics[width=0.32\textwidth]{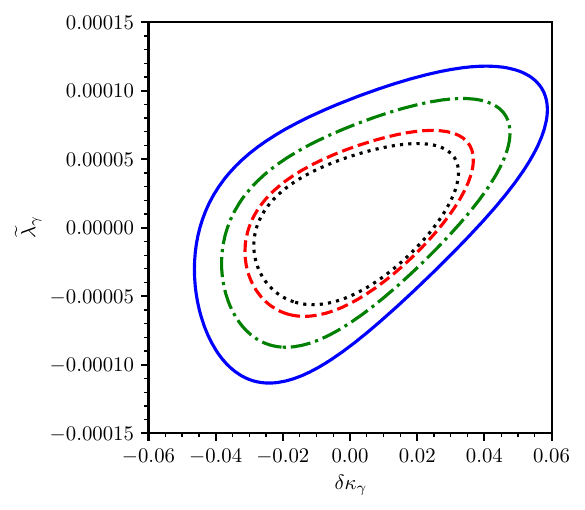}
    \includegraphics[width=0.32\textwidth]{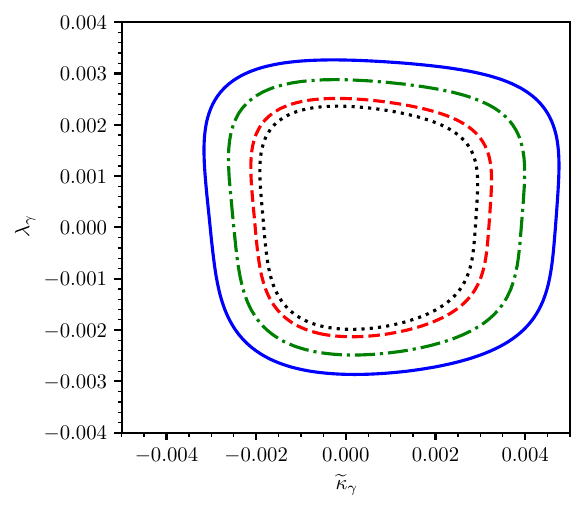}
    \includegraphics[width=0.32\textwidth]{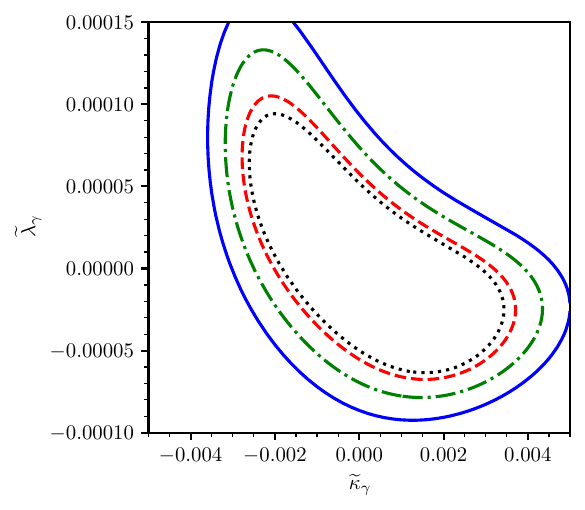}
    \includegraphics[width=0.32\textwidth]{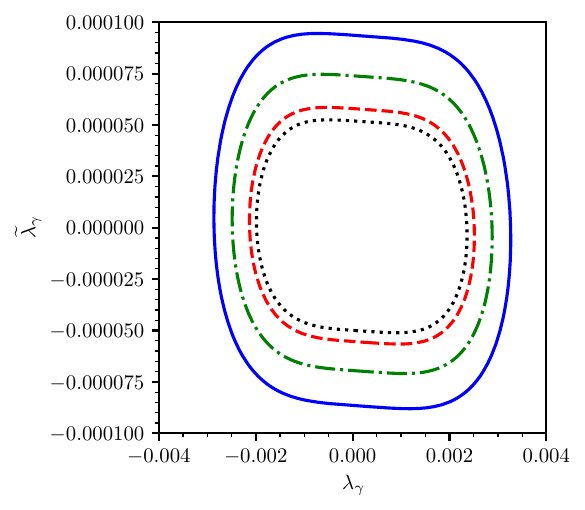}
    \caption{\label{fig:lumi}Two dimensional contours for asymmetries and $\sigma_{m_{W\gamma}}$ for four different values of integrated luminosities obtained by varying two aTGC simultaneously while keeping the other two to zero. The systematic errors for asymmetries are set to $\epsilon_{\mathcal{A}}=0.2\%$, and $\epsilon_\sigma=5\%$. }
\end{figure*}
\begin{figure*}[!htb]
    \centering
    \includegraphics[width=0.32\textwidth]{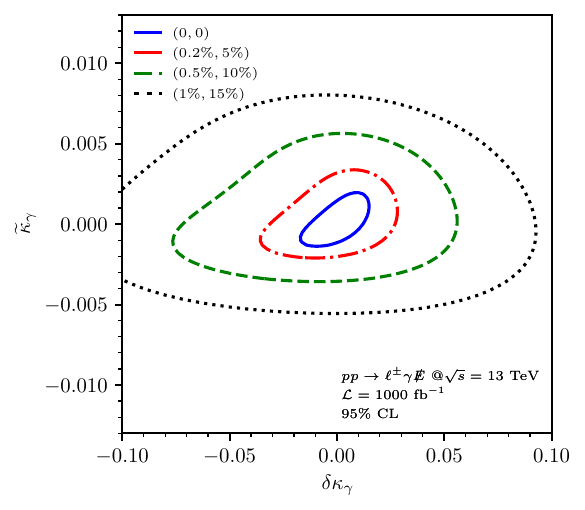}
    \includegraphics[width=0.32\textwidth]{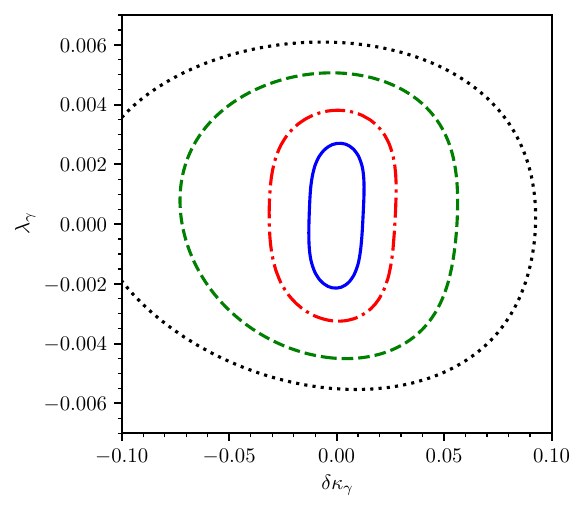}
    \includegraphics[width=0.32\textwidth]{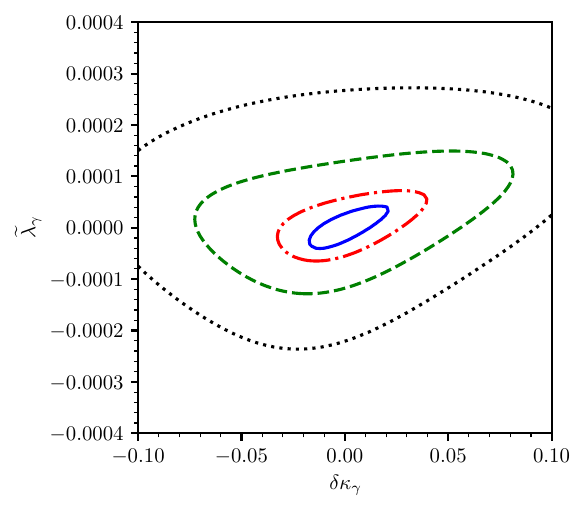}
    \includegraphics[width=0.32\textwidth]{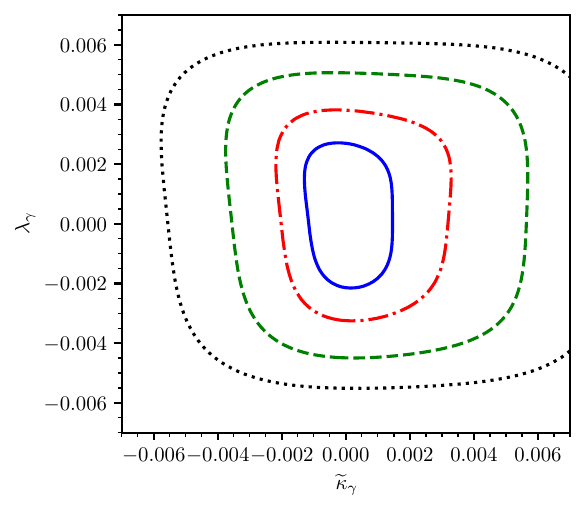}
    \includegraphics[width=0.32\textwidth]{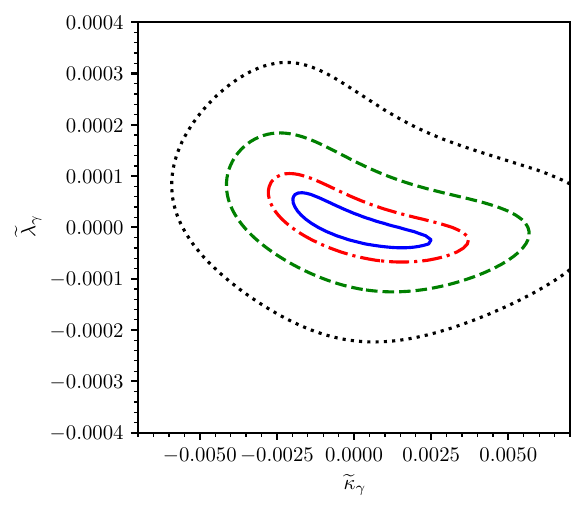}
    \includegraphics[width=0.32\textwidth]{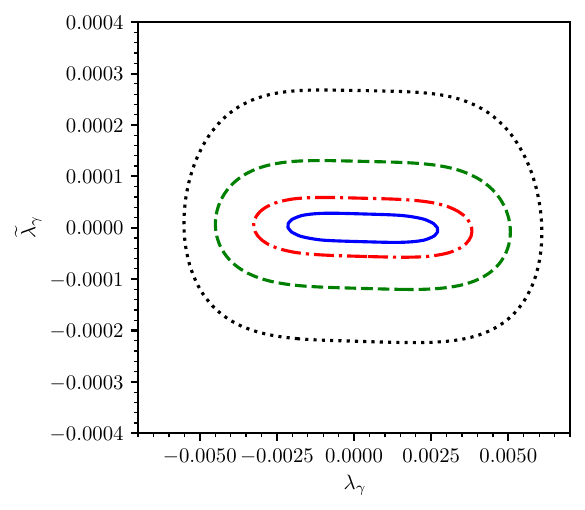}
    \caption{\label{fig:syst} Two dimensional contours of the asymmetries for four different values of systematic errors and $\mathcal{L}=1000$ fb$^{-1}$ obtained by varying two aTGC simultaneously while keeping the other two to zero. }
\end{figure*}

\begin{table*}[!htb]
    \centering
    \renewcommand{\arraystretch}{1.5}
    \caption{One parameter $95\%$ confidence level bounds on anomalous triple gauge couplings obtained by combining both asymmetries and $m_{W\gamma}$. The upper table tabulates the limits at four different values of integrated luminosities with systematic errors of $\epsilon_{\sigma}=5\%$ and $\epsilon_{\mathcal{A}}=0.2\%$, and the lower table lists the limits at $\mathcal{L}=1000$ fb$^{-1}$ for four different values of systematic errors.}
    \label{tab:lim95}
    \begin{tabular*}{0.99\textwidth}{@{\extracolsep{\fill}}ccccc@{}}\hline
     Params/Lumi $(\mathcal{L})$ & $140$ fb$^{-1}$ & $300$ fb$^{-1}$ & $1000$ fb$^{-1}$ & $3000$ fb$^{-1}$ \\ \hline
     $\delta\kappa_\gamma$ & $[-0.036,\,+0.031]$ & $[-0.030,\,+0.026]$ & $[-0.024,\,+0.021]$ & $[-0.022,\,+0.020]$ \\
     $\lambda_\gamma$ & $[-0.0034,+0.0038]$ & $[-0.0030,\,+0.0034]$ & $[-0.0026,\,+0.0030]$ & $[-0.0024,\,+0.0028]$ \\
     $\widetilde{\kappa}_\gamma$ & $[-0.0025,\,+0.0039]$ & $[-0.0021,+0.0032]$ & $[-0.0017,+0.0026]$ & $[-0.0015,\,+0.0023]$ \\
     $\widetilde{\lambda}_\gamma$ & $[-0.00007,\,+0.00007]$ & $[-0.00006,\,+0.00006]$ & $[-0.00004,\,+0.00005]$ & $[-0.00004,\,+0.00004]$ \\
     \hline
     Params/Syst ($\epsilon_A$)& $(0,0)$ & $(0.2\%,5\%)$ & $(0.5\%,10\%)$&$(1\%,15\%)$ \\
     \hline
     $\delta\kappa_\gamma$ & $[-0.010,\,+0.010]$ & $[-0.024,\,+0.021]$& $[-0.056,\,+0.046]$ & $[-0.090,\,+0.075]$ \\
     $\lambda_\gamma$ & $[-0.0019,\,+0.0023]$ & $[-0.0026,\,+0.0030]$& $[-0.0035,\,+0.0039]$ & $[-0.0042,\,+0.0046]$ \\
     $\widetilde{\kappa}_\gamma$ & $[-0.0010,\,+0.0011]$ & $[-0.0017,\,+0.0026]$ & $[-0.0030,\,+0.0049]$ & $[-0.0048,\,+0.0072]$ \\
     $\widetilde{\lambda}_\gamma$ & $[-0.00002,\,+0.00002]$ &$[-0.00004,\,+0.00005]$ &$[-0.0001,\,+0.0001]$ & $[-0.00018,+0.00021]$\\
     \hline
    \end{tabular*}
\end{table*}

Next, we estimate the impact of integrated luminosity on the bounds. In Fig.~\ref{fig:lumi}, we present $95\%$ CL $\Delta\chi^2$ contours for all six combinations of aTGC by combining the contribution from all eight asymmetries in all four regions of $\Delta\eta_{\ell\gamma}$ and $\sigma_{m_{W\gamma}}$ for different values of $\mathcal{L}$. The contours are obtained by varying two aTGCs at a time for four different benchmark values of $\mathcal{L}$
\begin{align}
    \mathcal{L} \in \{140,\,300,\,1000,\,3000\}~\mathrm{fb}^{-1}
\end{align}
with flat systematic errors of $\epsilon_{\mathcal{A}}=0.2\%$ and $\epsilon_\sigma=5\%$. The two pairs viz. $\delta\kappa_\gamma-\widetilde{\kappa}_\gamma$ and $\delta\kappa_\gamma-\widetilde{\lambda}_\gamma$ show a mild positive correlation, while $\widetilde{\kappa}_\gamma-\widetilde{\lambda}_\gamma$ has a negative correlation. The remaining three combinations present almost zero correlation.  The linear extent of all six contours shrinks modestly on increasing the luminosity from $140$ fb$^{-1}$ to $3000$ fb$^{-1}$ by a factor $\sim 1.4-1.8$ while the statistical uncertainty lowers by a factor of $4.6$. 

Fig.~\ref{fig:syst} shows the contours for all six combinations at fixed luminosity $\mathcal{L} = 1000~\text{fb}^{-1}$ and four different systematic uncertainty scenarios for asymmetries
\begin{align}
(\epsilon_{\mathcal{A}},\epsilon_\sigma) &= \{(0,0),\, (0.2\%,5\%), \, (0.5\%,10\%),\,(1\%,15\%)\}.
\end{align}
All contours show a significant increase in linear extent on raising systematics from an ideal zero case to a conservative value of $(1\%, 15\%)$, by factors ranging from $\sim 2$ ($\lambda_\gamma$) to $\sim 10$ ($\widetilde{\lambda}_\gamma$). The corresponding one-parameter $95\%$ CL limits on each aTGC are summarized in Table~\ref{tab:lim95} for both the luminosity and the systematic uncertainty scan.

The one parameter $95\%$ CL bounds in Table~\ref{tab:lim95} quantify the interplay between statistics and systematics. Increasing $\mathcal{L}$ from $140$ to $3000~\text{fb}^{-1}$, a factor of about $21$, improves the limits only by a factor of $\sim 1.4-1.8$, far below the $\sqrt{\mathcal{L}}$ scaling of $\sim 4.6$ expected in a statistics limited regime. This already indicates that systematic uncertainties dominate the sensitivity. Consistently, at $\mathcal{L}=1000~\text{fb}^{-1}$, moving the systematic scan from the ideal case to a conservative $(\epsilon_{\mathcal{A}},\epsilon_{\sigma})=(1\%,15\%)$ degrades the bounds by $\sim 8$ for $\delta\kappa_\gamma$, $\sim 2$ for $\lambda_\gamma$, $\sim 6$ for $\widetilde{\kappa}_\gamma$, and $\sim 10$ for $\widetilde{\lambda}_\gamma$. The variation of the bounds suggests that systematic errors, especially those entering the asymmetries, must be brought down to the percent level or below before additional luminosity can translate into improved precision; otherwise, increasing $\mathcal{L}$ alone yields only marginal gains in aTGC sensitivity.

On the experimental side, the tightest bound on the anomalous $W^-W^+\gamma$ has been reported in the SM effective field theory (SMEFT) framework in the HISZ basis by CMS~\cite{CMS:2026lik} in the combined $WW$ and $WZ$ analysis. One can match the limits in HISZ basis to that of the LEP parameters as~\cite{Degrande:2012wf}
\begin{align}
    \delta\kappa_\gamma &= (c_W + c_B)\,\frac{m_W^2}{2\,\Lambda^2},\nonumber\\
    \lambda_\gamma &= c_{WWW}\,\frac{3\,g_W^2\,m_W^2}{2\,\Lambda^2},\nonumber\\
    \widetilde{\kappa}_\gamma &= c_{\widetilde{W}}\,\tan^2\theta_W\,\frac{m_W^2}{2\,\Lambda^2},\nonumber\\
    \widetilde{\lambda}_\gamma &= c_{\widetilde{W}WW}\,\frac{3g_W^2m_W^2}{2\,\Lambda^2},
\end{align}
where $c_i$ are the Wilson coefficients (WCs) of the dimension-6 operators viz. $\mathscr{O}_W$, $\mathscr{O}_B$, $\mathscr{O}_{\widetilde{W}}$ and $\mathscr{O}_{\widetilde{W}WW}$ (see Ref.~\cite{Hagiwara:1993ck} for details). The observed $95\%$ CL bounds on the WCs at $\mathcal{L} = 138$ fb$^{-1}$ and the corresponding anomalous parameters are listed in Table~\ref{tab:exLim}.
\begin{table}[!h]
    \centering
    \caption{One parameter $95\%$ CL experimental bounds on the Wilson coefficients and the relevant anomalous couplings. The bounds are reported at an integrated luminosity $\mathcal{L}=138$~fb$^{-1}$ with combined $WW$ and $WZ$ final state by CMS~\cite{CMS:2026lik}, with $\Lambda=1$~TeV.}
    \label{tab:exLim}
    \renewcommand{\arraystretch}{1.3}
     \begin{tabular*}{\linewidth}{@{\extracolsep{\fill}}cccc@{}}\hline
     WCs & Limits (TeV$^{-2}$) & aTGCs & Limits  \\
     \hline
     $c_B/\Lambda^2$ & $[-5.18,+5.39]$ & & \\
     $c_W/\Lambda^2$ & $[-1.44,+1.78]$ &  $\delta\kappa_\gamma$ &$[-0.021,+0.023]$ \\
     $c_{WWW}/\Lambda^2$ & $[-0.93,+0.93]$ & $\lambda_\gamma$ & $[-0.004,+0.004]$ \\
     $c_{\widetilde{W}}/\Lambda^2$ & $[-1.31,+1.24]$ & $\widetilde{\kappa}_\gamma$ & $[-0.001,+0.001]$\\
     $c_{\widetilde{W}WW}/\Lambda^2$ & $[-0.37,+0.49]$ & $\widetilde{\lambda}_\gamma$ & $[-0.0015,+0.002]$ \\
    \hline
    \end{tabular*}
\end{table}

We compare our limits obtained at $140$ fb$^{-1}$ with $(\epsilon_A, \epsilon_\sigma) = (0.2\%,5\%)$ and $\Lambda=1$~TeV with those of CMS~\cite{CMS:2026lik} listed in Table~\ref{tab:exLim}. The CMS bound on the CP-even coupling $\delta\kappa_\gamma$ is tighter than ours by a factor of $\sim 1.5$. For $\lambda_\gamma$, the two limits are essentially comparable, with ours tighter by about $10\%$. For the CP-odd coupling $\widetilde{\kappa}_\gamma$, the CMS bound is stronger than ours by a factor of $\sim 3$. The most distinctive difference occurs for $\widetilde{\lambda}_\gamma$, where our bound is tighter than the CMS one by roughly a factor of $\sim 25$, reflecting the dedicated sensitivity of the polarization asymmetries to the CP-odd triple-gauge structure. 

\begin{figure*}[!t]
    \centering
    \includegraphics[width=0.325\linewidth]{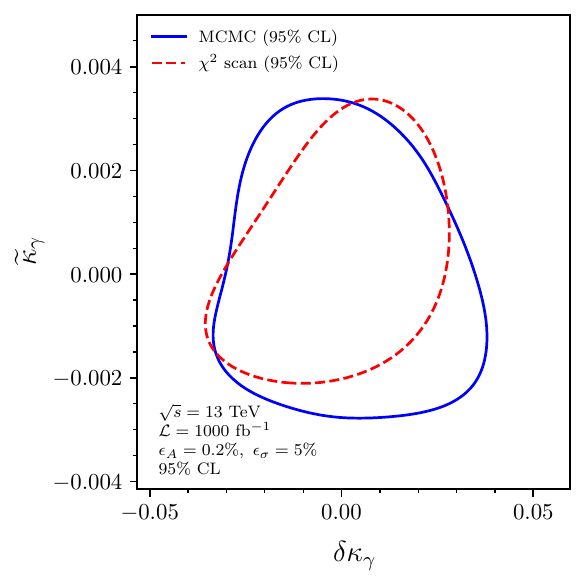}
    \includegraphics[width=0.325\linewidth]{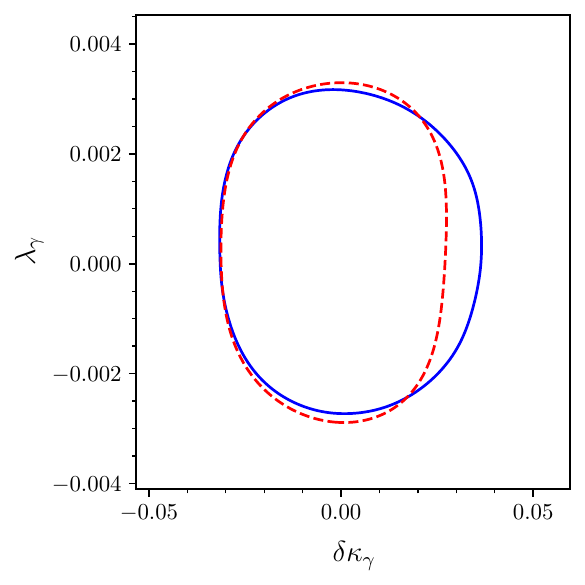}
    \includegraphics[width=0.325\linewidth]{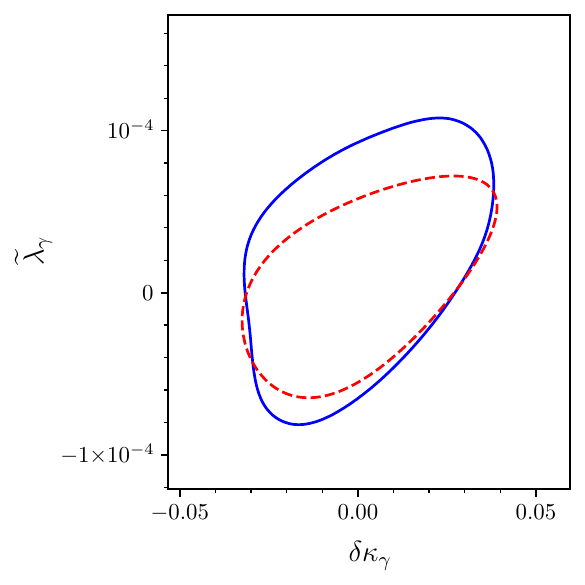}
    \includegraphics[width=0.325\linewidth]{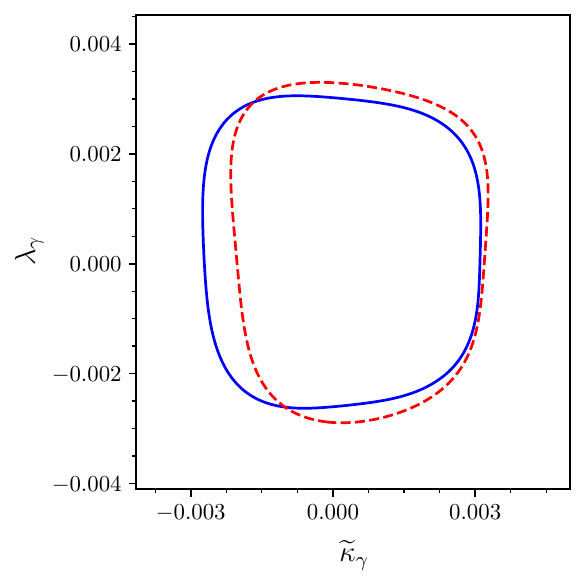}
    \includegraphics[width=0.325\linewidth]{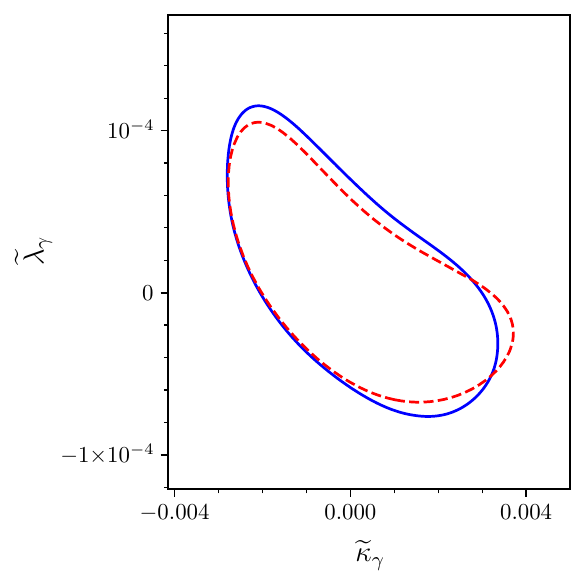}
    \includegraphics[width=0.325\linewidth]{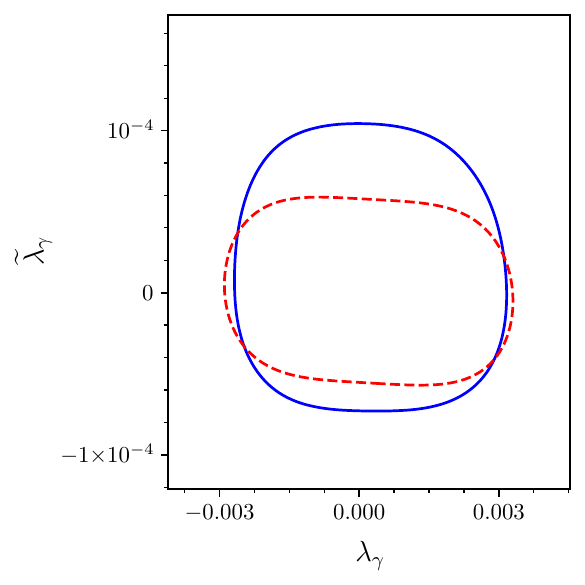}
    \caption{Two parameter $95\%$ CL contours for all six pairs of anomalous couplings obtained via a global fit with MCMC (solid blue curve) and chi-squared scan (dashed red curve). The analysis is performed for $pp\to \ell^\pm \gamma\slashed{E}$ at NLO QCD at $\sqrt{s}=13$ TeV, $\mathcal{L}=1000$ fb$^{-1}$ and $(\epsilon_A,\epsilon_\sigma) = (0.2\%,5\%)$.}
    \label{fig:mcmc}
\end{figure*}
\section{MCMC analysis}
\label{sec:mcmc}
The one and two parameter at a time analysis does not present a complete picture of anomalous couplings. One has to compute the behavior in a scenario when all the anomalous couplings under consideration are non-zero. In the current work, we obtain a marginalized distribution of aTGCs using Markov-Chain-Monte-Carlo (MCMC) method. The Markov chains are constructed using the Metropolis-Hastings algorithm. The likelihood function is defined using the $\chi^2$ in Eq.\eqref{eq:chi} as
\begin{align}
    P(f_i|A) \propto \mathrm{Exp}\Big(\frac{-\chi^2(f_i)}{2}\Big),
\end{align}
where $f_i$ is a set of four anomalous couplings. We perform the MCMC analysis combining all eight asymmetries from all four disjoint regions of $\Delta\eta_{\ell\gamma}$ and $\sigma_{m_{W\gamma}}$ at $\mathcal{L}=1000$ fb$^{-1}$, $\epsilon_\sigma=5\%$ and $\epsilon_A=0.2\%$. We ran $50$ independent chains, each with $5\times 10^4$ steps and a burn-in of $500$ steps. The convergence of the samples is quantified via the Gelman-Rubin $\hat{R}$ statistic. For each of the anomalous couplings, we found the $\hat{R}$ close to unity. 

Fig.~\ref{fig:mcmc} compares the $95\%$ CL allowed regions in the two-parameter planes obtained from the MCMC analysis with those from the $\chi^2$ scan. For visualization, the dashed red curve in Figs.~\ref{fig:lumi}, \ref{fig:syst} and \ref{fig:mcmc}, corresponds to the same run scenario.  Both agree well in the location and overall shape of the contours. A visible dilution of the MCMC region is present for most couplings, most notably in the $(\delta\kappa_\gamma$-$\widetilde{\kappa}_\gamma)$, $(\delta\kappa_\gamma$-$\widetilde{\lambda}_\gamma)$ and $(\widetilde{\kappa}_\gamma$-$\widetilde{\lambda}_\gamma)$ planes. In contrast, no such dilution appears along the $\lambda_\gamma$ axis: the MCMC limit on $\lambda_\gamma$ is comparable to, and marginally tighter than, the $\chi^2$ one. Table~\ref{tab:lim_mcmc} lists the marginalized one-parameter $95\%$ CL limits from the MCMC analysis. Apart from $\lambda_\gamma$, the MCMC bounds are systematically weaker than those from the $\chi^2$ scan by factors of up to $\sim 1.7$.

\begin{table}[!htb]
    \centering
    \caption{\label{tab:lim_mcmc}List of projected $95\%$ CL of anomalous triple gauge couplings from MCMC global fits at $\mathcal{L}=1000$ fb$^{-1}$ and systematic errors of $\epsilon_A=0.2\%$ and $\epsilon_\sigma=5\%$.}
    \renewcommand{\arraystretch}{1.4}
    \begin{tabular}{cc}
    \hline
     aTGCs & Limits  \\
     \hline
      $\delta\kappa_\gamma$   & $[-0.026,\,+0.031]$ \\
      $\lambda_\gamma$ & $[-0.0022,\,+0.0026]$ \\
      $\widetilde{\kappa}_\gamma$ & $[-0.0024,\,+0.0026]$ \\
      $\widetilde{\lambda}_\gamma$ & $[-0.00006,\,+0.00009]$ \\
      \hline
    \end{tabular}
\end{table}
\section{Electric dipole moments}
\label{sec:edm}
The presence of anomalous CP-violating $W^-W^+\gamma$ coupling could induce additional contribution to the fermionic electric dipole moment (EDM)~\cite{Novales-Sanchez:2007rsw,Boudjema:1990dv}. Further, the EDMs are strongly bounded from the low energy experiments. The current stringent bound on electron EDM (eEDM) at $90\%$ CL has been reported by the JILA HfF$^{+}$ experiment to be~\cite{Roussy:2022cmp}
\begin{equation}
    \frac{d_e}{e} <  4.1 \times 10^{-30}\,\mathrm{cm} \equiv 2.081\times 10^{-16}\, \mathrm{GeV}^{-1}.
\end{equation}
The contribution to the eEDM in presence of $\widetilde{\lambda}_\gamma$ and $\widetilde{\kappa}_\gamma$ at the one-loop level is given by~\cite{Panico:2018hal}
\begin{align}
    \frac{d_e}{e} &= \frac{\,y_e\,v}{16\pi^2\,g_W\,m_W^2}\,\widetilde{\lambda}_\gamma +\frac{\sqrt{2}}{16\pi^2}\frac{y_e}{v}\frac{\cos\theta_W}{\sin\theta_W}\Big[\sin(2\theta_W)\nonumber\\& +\frac{(1-2\sin^2\theta_W)(1-4\sin^2\theta_W)}{\sin(2\theta_W)} -\frac{\cos\theta_W}{2\,\sin\theta_W}\Big]\nonumber\\
    &\quad \times \mathrm{ln}\Big(\frac{\Lambda^2}{m_H^2}\Big)\,\widetilde{\kappa}_\gamma,
\end{align}
where $y_e = \sqrt{2}\,m_e/v \equiv 2.9\times 10^{-6}$ is the electron-Yukawa coupling, $m_H$ is the mass of the Higgs boson, and $\Lambda$ is the new physics scale. We set $g_W = 0.65$, $m_H = 125$~GeV, and $\Lambda=5$~TeV. With the above bound on the eEDM, we obtain the bounds on the CP-violating $W^-W^+\gamma$ couplings as
\begin{align}
    |\widetilde{\kappa}_\gamma| < 6.4\times 10^{-6}, \quad |\widetilde{\lambda}_\gamma| < 1.93\times 10^{-7}.
\end{align}
In comparison to our collider bounds at $\mathcal{L}=1000~\text{fb}^{-1}$ with $\epsilon_{\mathcal{A}}=0.2\%$ and $\epsilon_{\sigma}=5\%$, the low-energy eEDM measurement constrains both $\widetilde{\kappa}_\gamma$ and $\widetilde{\lambda}_\gamma$ by about $2.5$ orders of magnitude more tightly.  Using the above eEDM bounds, the electric dipole and magnetic quadrupole moments of the $W$ boson are constrained to $|d_W| < 1.2\times 10^{-8}~\text{GeV}^{-1}$ and $|\widetilde{Q}_W| < 3.1\times 10^{-10}~\text{GeV}^{-2}$, respectively.

\section{Conclusion}
\label{sec:con}

We have investigated the sensitivity of the $pp\to\ell^\pm\gamma\slashed{E}$ process to anomalous $W^-W^+\gamma$ interactions at NLO QCD, exploiting the polarization structure of the reconstructed $W$ boson. Our analysis combines the eight independent polarization asymmetries, measured in four disjoint regions of $\Delta\eta_{\ell\gamma}$, with the cross section in the high-energy tail of the reconstructed $m_{W\gamma} \ge 1$ TeV distribution. This provides complementary sensitivity to the four independent anomalous couplings, comprising two CP-even and two CP-odd interactions.

Our results demonstrate that the polarization observables probe different aspects of the anomalous vertex. In particular, the $\cos\theta_\ell$ asymmetry provides the dominant sensitivity to $\delta\kappa_\gamma$ and the CP-odd couplings, whereas the $\sigma_{m_{W\gamma}}$ is the principal probe of $\lambda_\gamma$. The sensitivity does not improve proportionally with the accumulated luminosity: increasing $\mathcal{L}$ from $140$ to $3000~\mathrm{fb}^{-1}$ strengthens the one-parameter $95\%$ CL bounds by only a factor of $\sim 1.4-1.8$, compared with the statistical expectation of $\sim 4.6$. The substantial degradation of the bounds when systematic uncertainties are increased further establishes that the projected sensitivity is systematics-dominated.

The simultaneous four-parameter MCMC analysis yields marginalized constraints consistent with the corresponding $\chi^2$ analysis, with differences of up to a factor of $\sim 1.7$. A comparison with the CMS $WW+WZ$ combination shows that our constraint with $\mathcal{L}=140$ fb$^{-1}$ and systematics $(\epsilon_A,\epsilon_\sigma)$ = $(0.2\%,5\%)$  is stronger by approximately a factor of $25$ on $\widetilde{\lambda}_\gamma$. Our bound on $\lambda_\gamma$ is comparable to the experimental one. Although the CMS bounds remain stronger for $\delta\kappa_\gamma$ and $\widetilde{\kappa}_\gamma$, the pronounced improvement in $\widetilde{\lambda}_\gamma$ highlights the sensitivity of polarization measurements to CP-odd structures of the $W^-W^+\gamma$ vertex. This establishes the $W\gamma$ channel as a complementary probe to conventional diboson measurements, with particular sensitivity to the CP-odd sector.

Finally, the electron EDM constraint provides a substantially stronger probe of CP violation, restricting $\widetilde{\kappa}_\gamma$ and $\widetilde{\lambda}_\gamma$ by approximately $2.5$ orders of magnitude beyond the collider reach. Nevertheless, collider measurements probe these interactions directly at the electroweak scale and provide complementary information on their contributions to the $W$-boson electromagnetic moments. Further improvements in the collider sensitivity will therefore depend critically on controlling systematic uncertainties. Combining polarization information from $W\gamma$ with spin-sensitive observables in other diboson channels, such as $WW$ and $WZ$, offers a promising direction for more comprehensive constraints on anomalous triple-gauge interactions.

\section*{Acknowledgements}
 Yu Shi and Amir Subba acknowledge the support of National Natural Science Foundation of China under Grant Nos. T2241005 and 12075059.

\bibliography{refer}

@article{CDF:2004tuf,
    author = "Acosta, D. and others",
    collaboration = "CDF",
    title = "{Measurement of $W \gamma$ and $Z \gamma$ production in $p\bar{p}$ collisions at $\sqrt{s} = 1.96$ TeV}",
    eprint = "hep-ex/0410008",
    archivePrefix = "arXiv",
    reportNumber = "FERMILAB-PUB-04-246-E",
    doi = "10.1103/PhysRevLett.94.041803",
    journal = "Phys. Rev. Lett.",
    volume = "94",
    pages = "041803",
    year = "2005"
}

@article{D0:2005djn,
    author = "Abazov, V. M. and others",
    collaboration = "D0",
    title = "{Measurement of the $p$ - $\bar{p} \to W \gamma$ + $X$ cross section at $\sqrt{s}$ = 1.96-TeV and $W W \gamma$ anomalous coupling limits}",
    eprint = "hep-ex/0503048",
    archivePrefix = "arXiv",
    reportNumber = "FERMILAB-PUB-05-046-E",
    doi = "10.1103/PhysRevD.71.091108",
    journal = "Phys. Rev. D",
    volume = "71",
    pages = "091108",
    year = "2005"
}

@article{D0:2008abl,
    author = "Abazov, V. M. and others",
    collaboration = "D0",
    title = "{First study of the radiation-amplitude zero in $W \gamma$ production and limits on anomalous $W W \gamma$ couplings at $\sqrt{s}$ = 1.96- TeV}",
    eprint = "0803.0030",
    archivePrefix = "arXiv",
    primaryClass = "hep-ex",
    reportNumber = "FERMILAB-PUB-08-052-E",
    doi = "10.1103/PhysRevLett.100.241805",
    journal = "Phys. Rev. Lett.",
    volume = "100",
    pages = "241805",
    year = "2008"
}

@article{Brown:1979ux,
    author = "Brown, R. W. and Sahdev, D. and Mikaelian, K. O.",
    title = "{W+- Z0 and W+- gamma Pair Production in Neutrino e, p p, and anti-p p Collisions}",
    reportNumber = "Print-79-0404 (CASE WESTERN)",
    doi = "10.1103/PhysRevD.20.1164",
    journal = "Phys. Rev. D",
    volume = "20",
    pages = "1164",
    year = "1979"
}

@article{Baur:1988qt,
    author = "Baur, U. and Zeppenfeld, D.",
    title = "{Probing the W W gamma Vertex at Future Hadron Colliders}",
    reportNumber = "CERN-TH-5006/88, MAD/PH/412",
    doi = "10.1016/0550-3213(88)90045-4",
    journal = "Nucl. Phys. B",
    volume = "308",
    pages = "127--148",
    year = "1988"
}

@article{Baur:1989gk,
    author = "Baur, U. and Berger, Edmond L.",
    title = "{Probing the $W W \gamma$ Vertex at the Tevatron Collider}",
    reportNumber = "CERN-TH-5517/89, ANL-HEP-PR-89-86",
    doi = "10.1103/PhysRevD.41.1476",
    journal = "Phys. Rev. D",
    volume = "41",
    pages = "1476",
    year = "1990"
}

@article{Mikaelian:1979nr,
    author = "Mikaelian, K. O. and Samuel, M. A. and Sahdev, D.",
    title = "{The Magnetic Moment of Weak Bosons Produced in p p and p anti-p Collisions}",
    reportNumber = "PRINT-79-0454 (OKLAHOMA-STATE)",
    doi = "10.1103/PhysRevLett.43.746",
    journal = "Phys. Rev. Lett.",
    volume = "43",
    pages = "746",
    year = "1979"
}

@article{Goebel:1980es,
    author = "Goebel, C. J. and Halzen, F. and Leveille, J. P.",
    title = "{Angular zeros of Brown, Mikaelian, Sahdev, and Samuel and the factorization of tree amplitudes in gauge theories}",
    reportNumber = "DOE-ER/00881-172",
    doi = "10.1103/PhysRevD.23.2682",
    journal = "Phys. Rev. D",
    volume = "23",
    pages = "2682--2685",
    year = "1981"
}

@article{Brodsky:1982sh,
    author = "Brodsky, Stanley J. and Brown, Robert W.",
    title = "{Zeros in Amplitudes: Gauge Theory and Radiation Interference}",
    reportNumber = "FERMILAB-PUB-82-035-THY, FERMILAB-PUB-82-035-T",
    doi = "10.1103/PhysRevLett.49.966",
    journal = "Phys. Rev. Lett.",
    volume = "49",
    pages = "966",
    year = "1982"
}

@article{Brown:1982xx,
    author = "Brown, Robert W. and Kowalski, K. L. and Brodsky, Stanley J.",
    title = "{Classical Radiation Zeros in Gauge Theory Amplitudes}",
    reportNumber = "SLAC-REPRINT-1983-004, FERMILAB-PUB-82-102-T, FERMILAB-PUB-82-102-THY, CWRUTH-83-8",
    doi = "10.1103/PhysRevD.29.2100",
    journal = "Phys. Rev. D",
    volume = "28",
    pages = "624",
    year = "1983",
    note = "[Addendum: Phys.Rev.D 29, 2100--2104 (1984)]"
}

@article{CMS:2011myh,
    author = "Chatrchyan, Serguei and others",
    collaboration = "CMS",
    title = "{Measurement of $W\gamma$ and $Z\gamma$ production in $pp$ collisions at $\sqrt{s} = 7$ TeV}",
    eprint = "1105.2758",
    archivePrefix = "arXiv",
    primaryClass = "hep-ex",
    reportNumber = "CERN-PH-EP-2011-045, CMS-EWK-10-008",
    doi = "10.1016/j.physletb.2011.06.034",
    journal = "Phys. Lett. B",
    volume = "701",
    pages = "535--555",
    year = "2011"
}

@article{CMS:2013ryd,
    author = "Chatrchyan, Serguei and others",
    collaboration = "CMS",
    title = "{Measurement of the $W\gamma$ and $Z\gamma$ Inclusive Cross Sections in $pp$ Collisions at $\sqrt s=7$  TeV and Limits on Anomalous Triple Gauge Boson Couplings}",
    eprint = "1308.6832",
    archivePrefix = "arXiv",
    primaryClass = "hep-ex",
    reportNumber = "CMS-EWK-11-009, CERN-PH-EP-2013-108",
    doi = "10.1103/PhysRevD.89.092005",
    journal = "Phys. Rev. D",
    volume = "89",
    number = "9",
    pages = "092005",
    year = "2014"
}

@article{CMS:2021foa,
    author = "Sirunyan, Albert M and others",
    collaboration = "CMS",
    title = "{Measurement of the W$\gamma$ Production Cross Section in Proton-Proton Collisions at $\sqrt {s}$=13{\,}{\,}TeV and Constraints on Effective Field Theory Coefficients}",
    eprint = "2102.02283",
    archivePrefix = "arXiv",
    primaryClass = "hep-ex",
    reportNumber = "CMS-SMP-19-002, CERN-EP-2021-008",
    doi = "10.1103/PhysRevLett.126.252002",
    journal = "Phys. Rev. Lett.",
    volume = "126",
    number = "25",
    pages = "252002",
    year = "2021"
}

@article{CMS:2021cxr,
    author = "Tumasyan, Armen and others",
    collaboration = "CMS",
    title = "{Measurement of W$^\pm\gamma$ differential cross sections in proton-proton collisions at $\sqrt{s}$ = 13 TeV and effective field theory constraints}",
    eprint = "2111.13948",
    archivePrefix = "arXiv",
    primaryClass = "hep-ex",
    reportNumber = "CMS-SMP-20-005, CERN-EP-2021-219",
    doi = "10.1103/PhysRevD.105.052003",
    journal = "Phys. Rev. D",
    volume = "105",
    number = "5",
    pages = "052003",
    year = "2022"
}

@article{ATLAS:2012bpb,
    author = "Aad, Georges and others",
    collaboration = "ATLAS",
    title = "{Measurement of $W \gamma$ and $Z \gamma$ production cross sections in $pp$ collisions at $\sqrt{s}=7$ TeV and limits on anomalous triple gauge couplings with the ATLAS detector}",
    eprint = "1205.2531",
    archivePrefix = "arXiv",
    primaryClass = "hep-ex",
    reportNumber = "CERN-PH-EP-2012-059",
    doi = "10.1016/j.physletb.2012.09.017",
    journal = "Phys. Lett. B",
    volume = "717",
    pages = "49--69",
    year = "2012"
}

@article{ATLAS:2012upi,
    author = "Aad, Georges and others",
    collaboration = "ATLAS",
    title = "{Measurement of the $W W$ cross section in $\sqrt{s}=7$ TeV $pp$ collisions with the ATLAS detector and limits on anomalous gauge couplings}",
    eprint = "1203.6232",
    archivePrefix = "arXiv",
    primaryClass = "hep-ex",
    reportNumber = "CERN-PH-EP-2012-060",
    doi = "10.1016/j.physletb.2012.05.003",
    journal = "Phys. Lett. B",
    volume = "712",
    pages = "289--308",
    year = "2012"
}

@article{ATLAS:2013way,
    author = "Aad, Georges and others",
    collaboration = "ATLAS",
    title = "{Measurements of $W \gamma$ and $Z \gamma$ production in $pp$ collisions at $\sqrt{s}$=7  TeV with the ATLAS detector at the LHC}",
    eprint = "1302.1283",
    archivePrefix = "arXiv",
    primaryClass = "hep-ex",
    reportNumber = "CERN-PH-EP-2012-345",
    doi = "10.1103/PhysRevD.87.112003",
    journal = "Phys. Rev. D",
    volume = "87",
    number = "11",
    pages = "112003",
    year = "2013",
    note = "[Erratum: Phys.Rev.D 91, 119901 (2015)]"
}

@article{Boudjema:2009fz,
    author = "Boudjema, Fawzi and Singh, Ritesh K.",
    title = "{A Model independent spin analysis of fundamental particles using azimuthal asymmetries}",
    eprint = "0903.4705",
    archivePrefix = "arXiv",
    primaryClass = "hep-ph",
    reportNumber = "LAPTH-1322",
    doi = "10.1088/1126-6708/2009/07/028",
    journal = "JHEP",
    volume = "07",
    pages = "028",
    year = "2009"
}

@article{Hagiwara:1986vm,
    author = "Hagiwara, Kaoru and Peccei, R. D. and Zeppenfeld, D. and Hikasa, K.",
    title = "{Probing the Weak Boson Sector in $e^+e^-\to W^+W^-$}",
    reportNumber = "MAD/PH/279, DESY-86-058",
    doi = "10.1016/0550-3213(87)90685-7",
    journal = "Nucl. Phys. B",
    volume = "282",
    pages = "253--307",
    year = "1987"
}

@article{Hagiwara:1993ck,
    author = "Hagiwara, Kaoru and Ishihara, S. and Szalapski, R. and Zeppenfeld, D.",
    title = "{Low-energy effects of new interactions in the electroweak boson sector}",
    reportNumber = "MAD-PH-737, UT-635, KEK-TH-356, KEK-PREPRINT-92-214",
    doi = "10.1103/PhysRevD.48.2182",
    journal = "Phys. Rev. D",
    volume = "48",
    pages = "2182--2203",
    year = "1993"
}

@article{Degrande:2012wf,
    author = "Degrande, Celine and Greiner, Nicolas and Kilian, Wolfgang and Mattelaer, Olivier and Mebane, Harrison and Stelzer, Tim and Willenbrock, Scott and Zhang, Cen",
    title = "{Effective Field Theory: A Modern Approach to Anomalous Couplings}",
    eprint = "1205.4231",
    archivePrefix = "arXiv",
    primaryClass = "hep-ph",
    reportNumber = "MPP-2011-149, SI-HEP-2011-17, CP3-12-25",
    doi = "10.1016/j.aop.2013.04.016",
    journal = "Annals Phys.",
    volume = "335",
    pages = "21--32",
    year = "2013"
}

@misc{Samuel:1997zs,
    author = "Samuel, Mark A and Abraha, Tesfaye",
    title = "{Finding the radiation amplitude zero in $W$ gamma production. Is it unique to the standard model?}",
    eprint = "hep-ph/9706336",
    archivePrefix = "arXiv",
    reportNumber = "SLAC-PUB-9762, OSU-RN-326",
    month = "6",
    year = "1997"
}

@article{Baur:1993ir,
    author = "Baur, U. and Han, Tao and Ohnemus, J.",
    title = "{QCD corrections to hadronic $W \gamma$ production with nonstandard $W W \gamma$ couplings}",
    eprint = "hep-ph/9305314",
    archivePrefix = "arXiv",
    reportNumber = "FSU-HEP-930519, FERMILAB-PUB-93-096-T, DTP-93-26",
    doi = "10.1103/PhysRevD.48.5140",
    journal = "Phys. Rev. D",
    volume = "48",
    pages = "5140--5161",
    year = "1993"
}

@article{Grazzini:2015nwa,
    author = "Grazzini, Massimiliano and Kallweit, Stefan and Rathlev, Dirk",
    title = "{$W\gamma$ and $Z\gamma$ production at the LHC in NNLO QCD}",
    eprint = "1504.01330",
    archivePrefix = "arXiv",
    primaryClass = "hep-ph",
    reportNumber = "ZU-TH-04-15, MITP-15-021",
    doi = "10.1007/JHEP07(2015)085",
    journal = "JHEP",
    volume = "07",
    pages = "085",
    year = "2015"
}

@article{Aguilar-Saavedra:2015yza,
    author = "Aguilar-Saavedra, J. A. and Bernabeu, J.",
    title = "{Breaking down the entire W boson spin observables from its decay}",
    eprint = "1508.04592",
    archivePrefix = "arXiv",
    primaryClass = "hep-ph",
    doi = "10.1103/PhysRevD.93.011301",
    journal = "Phys. Rev. D",
    volume = "93",
    number = "1",
    pages = "011301",
    year = "2016"
}

@article{Rahaman:2021fcz,
    author = "Rahaman, Rafiqul and Singh, Ritesh K.",
    title = "{Breaking down the entire spectrum of spin correlations of a pair of particles involving fermions and gauge bosons}",
    eprint = "2109.09345",
    archivePrefix = "arXiv",
    primaryClass = "hep-ph",
    reportNumber = "HRI-RECAPP-2021-010",
    doi = "10.1016/j.nuclphysb.2022.115984",
    journal = "Nucl. Phys. B",
    volume = "984",
    pages = "115984",
    year = "2022"
}

@article{Mirkes:1992hu,
    author = "Mirkes, E.",
    title = "{Angular decay distribution of leptons from W bosons at NLO in hadronic collisions}",
    reportNumber = "TTP-92-12",
    doi = "10.1016/0550-3213(92)90046-E",
    journal = "Nucl. Phys. B",
    volume = "387",
    pages = "3--85",
    year = "1992"
}

@article{Alwall:2014hca,
    author = "Alwall, J. and Frederix, R. and Frixione, S. and Hirschi, V. and Maltoni, F. and Mattelaer, O. and Shao, H. -S. and Stelzer, T. and Torrielli, P. and Zaro, M.",
    title = "{The automated computation of tree-level and next-to-leading order differential cross sections, and their matching to parton shower simulations}",
    eprint = "1405.0301",
    archivePrefix = "arXiv",
    primaryClass = "hep-ph",
    reportNumber = "CERN-PH-TH-2014-064, CP3-14-18, LPN14-066, MCNET-14-09, ZU-TH-14-14",
    doi = "10.1007/JHEP07(2014)079",
    journal = "JHEP",
    volume = "07",
    pages = "079",
    year = "2014"
}

@article{Alloul:2013bka,
    author = "Alloul, Adam and Christensen, Neil D. and Degrande, C{\'e}line and Duhr, Claude and Fuks, Benjamin",
    title = "{FeynRules  2.0 - A complete toolbox for tree-level phenomenology}",
    eprint = "1310.1921",
    archivePrefix = "arXiv",
    primaryClass = "hep-ph",
    reportNumber = "CERN-PH-TH-2013-239, MCNET-13-14, IPPP-13-71, DCPT-13-142, PITT-PACC-1308",
    doi = "10.1016/j.cpc.2014.04.012",
    journal = "Comput. Phys. Commun.",
    volume = "185",
    pages = "2250--2300",
    year = "2014"
}

@article{Degrande:2011ua,
    author = "Degrande, Celine and Duhr, Claude and Fuks, Benjamin and Grellscheid, David and Mattelaer, Olivier and Reiter, Thomas",
    title = "{UFO - The Universal FeynRules Output}",
    eprint = "1108.2040",
    archivePrefix = "arXiv",
    primaryClass = "hep-ph",
    reportNumber = "CP3-11-25, IPHC-PHENO-11-04, IPPP-11-39, DCPT-11-78, MPP-2011-68",
    doi = "10.1016/j.cpc.2012.01.022",
    journal = "Comput. Phys. Commun.",
    volume = "183",
    pages = "1201--1214",
    year = "2012"
}

@article{Darme:2023jdn,
    author = "Darm{\'e}, Luc and others",
    title = "{UFO 2.0: the {\textquoteleft}Universal Feynman Output{\textquoteright} format}",
    eprint = "2304.09883",
    archivePrefix = "arXiv",
    primaryClass = "hep-ph",
    reportNumber = "BONN-TH-2023-03, DESY-23-051, FERMILAB-PUB-23-138-T, KA-TP-06-2023,
  MCNET-23-06, P3H-23-023, TIF-UNIMI-2023-11",
    doi = "10.1140/epjc/s10052-023-11780-9",
    journal = "Eur. Phys. J. C",
    volume = "83",
    number = "7",
    pages = "631",
    year = "2023"
}

@article{Degrande:2014vpa,
    author = "Degrande, Celine",
    title = "{Automatic evaluation of UV and R2 terms for beyond the Standard Model Lagrangians: a proof-of-principle}",
    eprint = "1406.3030",
    archivePrefix = "arXiv",
    primaryClass = "hep-ph",
    reportNumber = "IPPP-14-58, DCPT-14-116",
    doi = "10.1016/j.cpc.2015.08.015",
    journal = "Comput. Phys. Commun.",
    volume = "197",
    pages = "239--262",
    year = "2015"
}

@article{Bierlich:2022pfr,
    author = "Bierlich, Christian and others",
    title = "{A comprehensive guide to the physics and usage of PYTHIA 8.3}",
    eprint = "2203.11601",
    archivePrefix = "arXiv",
    primaryClass = "hep-ph",
    reportNumber = "LU-TP 22-16, MCNET-22-04, FERMILAB-PUB-22-227-SCD",
    doi = "10.21468/SciPostPhysCodeb.8",
    journal = "SciPost Phys. Codeb.",
    volume = "2022",
    pages = "8",
    year = "2022"
}

@article{deFavereau:2013fsa,
    author = "de Favereau, J. and Delaere, C. and Demin, P. and Giammanco, A. and Lema{\^\i}tre, V. and Mertens, A. and Selvaggi, M.",
    collaboration = "DELPHES 3",
    title = "{DELPHES 3, A modular framework for fast simulation of a generic collider experiment}",
    eprint = "1307.6346",
    archivePrefix = "arXiv",
    primaryClass = "hep-ex",
    doi = "10.1007/JHEP02(2014)057",
    journal = "JHEP",
    volume = "02",
    pages = "057",
    year = "2014"
}

@article{Cacciari:2008gp,
    author = "Cacciari, Matteo and Salam, Gavin P. and Soyez, Gregory",
    title = "{The anti-$k_t$ jet clustering algorithm}",
    eprint = "0802.1189",
    archivePrefix = "arXiv",
    primaryClass = "hep-ph",
    reportNumber = "LPTHE-07-03",
    doi = "10.1088/1126-6708/2008/04/063",
    journal = "JHEP",
    volume = "04",
    pages = "063",
    year = "2008"
}

@article{Pelliccioli:2026ltl,
    author = "Pelliccioli, Giovanni and Re, Emanuele",
    title = "{SMEFT effects on spin correlations and entanglement at NLO QCD in diboson production at hadron colliders}",
    eprint = "2601.09540",
    archivePrefix = "arXiv",
    primaryClass = "hep-ph",
    reportNumber = "COMETA-2026-01, LAPTH-003/26",
    doi = "10.1103/mr8r-bvcd",
    journal = "Phys. Rev. D",
    volume = "113",
    number = "9",
    pages = "093005",
    year = "2026"
}

@article{Denner:2014bna,
    author = "Denner, Ansgar and Dittmaier, Stefan and Hecht, Markus and Pasold, Christian",
    title = "{NLO QCD and electroweak corrections to W+{\textbackslash}gamma{\textbackslash} production with leptonic W-boson decays}",
    eprint = "1412.7421",
    archivePrefix = "arXiv",
    primaryClass = "hep-ph",
    reportNumber = "FR-PHENO-2014-014",
    doi = "10.1007/JHEP04(2015)018",
    journal = "JHEP",
    volume = "04",
    pages = "018",
    year = "2015"
}

@article{Accomando:2005ra,
    author = "Accomando, E. and Denner, Ansgar and Meier, C.",
    title = "{Electroweak corrections to $W \gamma$ and $Z \gamma$ production at the LHC}",
    eprint = "hep-ph/0509234",
    archivePrefix = "arXiv",
    reportNumber = "PSI-PR-05-09, ZU-TH-18-05, DFTT-30-05",
    doi = "10.1140/epjc/s2006-02521-y",
    journal = "Eur. Phys. J. C",
    volume = "47",
    pages = "125--146",
    year = "2006"
}

@article{Accomando:2001fn,
    author = "Accomando, Elena and Denner, Ansgar and Pozzorini, Stefano",
    title = "{Electroweak correction effects in gauge boson pair production at the CERN LHC}",
    eprint = "hep-ph/0110114",
    archivePrefix = "arXiv",
    reportNumber = "PSI-PR-01-13, ZU-TH-35-01",
    doi = "10.1103/PhysRevD.65.073003",
    journal = "Phys. Rev. D",
    volume = "65",
    pages = "073003",
    year = "2002"
}

@article{Smith:1989xz,
    author = "Smith, J. and Thomas, D. and van Neerven, W. L.",
    title = "{{QCD} Corrections to the Reaction $p \bar{p} \to W \gamma$ X}",
    reportNumber = "CPT-89-P-2239",
    doi = "10.1007/BF01557332",
    journal = "Z. Phys. C",
    volume = "44",
    pages = "267",
    year = "1989"
}

@article{PhysRevD.47.940,
  title = {Order-${\ensuremath{\alpha}}_{s}$ calculations of hadronic ${W}^{\ifmmode\pm\else\textpm\fi{}}\ensuremath{\gamma}$ and $Z\ensuremath{\gamma}$ production},
  author = {Ohnemus, J.},
  journal = {Phys. Rev. D},
  volume = {47},
  issue = {3},
  pages = {940--955},
  numpages = {0},
  year = {1993},
  month = {Feb},
  publisher = {American Physical Society},
  doi = {10.1103/PhysRevD.47.940},
  url = {https://link.aps.org/doi/10.1103/PhysRevD.47.940}
}

@article{Dixon:1998py,
    author = "Dixon, Lance J. and Kunszt, Z. and Signer, A.",
    title = "{Helicity amplitudes for O(alpha-s) production of $W^{+} W^{-}$, $W^\pm Z$, $Z Z$, $W^\pm \gamma$, or $Z \gamma$ pairs at hadron colliders}",
    eprint = "hep-ph/9803250",
    archivePrefix = "arXiv",
    reportNumber = "SLAC-PUB-7750, ETH-TH-98-07, CERN-TH-98-66",
    doi = "10.1016/S0550-3213(98)00421-0",
    journal = "Nucl. Phys. B",
    volume = "531",
    pages = "3--23",
    year = "1998"
}

@article{Grazzini:2016hai,
    author = "Grazzini, Massimiliano and Kallweit, Stefan and Rathlev, Dirk",
    title = "{Wgamma and Zgamma production at the LHC in NNLO QCD}",
    eprint = "1601.06751",
    archivePrefix = "arXiv",
    primaryClass = "hep-ph",
    reportNumber = "MITP-16-010, DESY-16-015",
    doi = "10.22323/1.235.0074",
    journal = "PoS",
    volume = "RADCOR2015",
    pages = "074",
    year = "2016"
}

@article{Campbell:2021mlr,
    author = "Campbell, John M. and De Laurentis, Giuseppe and Ellis, R. Keith and Seth, Satyajit",
    title = "{The pp {\textrightarrow} W({\textrightarrow} l{\ensuremath{\nu}}) + {\ensuremath{\gamma}} process at next-to-next-to-leading order}",
    eprint = "2105.00954",
    archivePrefix = "arXiv",
    primaryClass = "hep-ph",
    reportNumber = "FERMILAB-PUB-21-215-T, FR-PHENO-2020-020, IPPP/20/96",
    doi = "10.1007/JHEP07(2021)079",
    journal = "JHEP",
    volume = "07",
    pages = "079",
    year = "2021"
}

@article{Etesami:2016rwu,
    author = "Etesami, Seyed Mohsen and Khatibi, Sara and Mohammadi Najafabadi, Mojtaba",
    title = "{Measuring anomalous WW $\gamma $ and t $\bar{\text {t}}\gamma $ couplings using top+ $\gamma $ production at the LHC}",
    eprint = "1606.02178",
    archivePrefix = "arXiv",
    primaryClass = "hep-ph",
    doi = "10.1140/epjc/s10052-016-4376-2",
    journal = "Eur. Phys. J. C",
    volume = "76",
    number = "10",
    pages = "533",
    year = "2016"
}

@article{Papavassiliou:1999iv,
    author = "Papavassiliou, Joannis and Philippides, Kostas",
    title = "{Probing the $W W \gamma$ vertex at hadron colliders}",
    eprint = "hep-ph/9907376",
    archivePrefix = "arXiv",
    reportNumber = "CERN-TH-99-212",
    doi = "10.1103/PhysRevD.60.113007",
    journal = "Phys. Rev. D",
    volume = "60",
    pages = "113007",
    year = "1999"
}

@article{He:1993hx,
    author = "He, Xiao-Gang and McKellar, Bruce",
    title = "{Constraints on the anomalous W W gamma couplings from b ---{\ensuremath{>}} s gamma}",
    eprint = "hep-ph/9309228",
    archivePrefix = "arXiv",
    reportNumber = "UM-P-93-52, OZ-93-14",
    doi = "10.1016/0370-2693(94)90841-9",
    journal = "Phys. Lett. B",
    volume = "320",
    pages = "165--169",
    year = "1994"
}

@article{Sahin:2017rfr,
    author = "{\c{S}}ahin, B.",
    title = "{Probing CP-violating $WW\gamma$ couplings in $\gamma p$ collision at the LHC}",
    doi = "10.1142/S0217732317502054",
    journal = "Mod. Phys. Lett. A",
    volume = "32",
    number = "37",
    pages = "1750205",
    year = "2017"
}

@article{Bhatia:2018ndx,
    author = "Bhatia, Disha and Maitra, Ushoshi and Raychaudhuri, Sreerup",
    title = "{Pinning down anomalous $WW\gamma$ couplings at the LHC}",
    eprint = "1804.05357",
    archivePrefix = "arXiv",
    primaryClass = "hep-ph",
    reportNumber = "TIFR-TH/18-9, TIFR-TH-18-9",
    doi = "10.1103/PhysRevD.99.095017",
    journal = "Phys. Rev. D",
    volume = "99",
    number = "9",
    pages = "095017",
    year = "2019"
}

@article{PhysRevD.51.90,
  title = {Hadron collider limits on anomalous WW\ensuremath{\gamma} couplings},
  author = {Barger, Kevin R. and Reno, M. H.},
  journal = {Phys. Rev. D},
  volume = {51},
  issue = {1},
  pages = {90--98},
  numpages = {0},
  year = {1995},
  month = {Jan},
  publisher = {American Physical Society},
  doi = {10.1103/PhysRevD.51.90},
  url = {https://link.aps.org/doi/10.1103/PhysRevD.51.90}
}

@article{Subba:2023rpm,
    author = "Subba, Amir and Singh, Ritesh K.",
    title = "{Study of anomalous $W^-W^+\gamma /Z$ couplings using polarizations and spin correlations in $e^-e^+\rightarrow W^-W^+$ with polarized beams}",
    eprint = "2305.15106",
    archivePrefix = "arXiv",
    primaryClass = "hep-ph",
    doi = "10.1140/epjc/s10052-023-12292-2",
    journal = "Eur. Phys. J. C",
    volume = "83",
    number = "12",
    pages = "1119",
    year = "2023"
}

@article{PhysRevD.107.073004,
  title = {Role of polarizations and spin-spin correlations of ${W}^{\ensuremath{'}}s$ in ${e}^{\ensuremath{-}}{e}^{+}\ensuremath{\rightarrow}{W}^{\ensuremath{-}}{W}^{+}$ at $\sqrt{s}=250\text{ }\text{ }\mathrm{GeV}$ to probe anomalous ${W}^{\ensuremath{-}}{W}^{+}Z/\ensuremath{\gamma}$ couplings},
  author = {Subba, Amir and Singh, Ritesh K.},
  journal = {Phys. Rev. D},
  volume = {107},
  issue = {7},
  pages = {073004},
  numpages = {18},
  year = {2023},
  month = {Apr},
  publisher = {American Physical Society},
  doi = {10.1103/PhysRevD.107.073004},
  url = {https://link.aps.org/doi/10.1103/PhysRevD.107.073004}
}

@article{CHOUDHURY1999193,
title = {CP-violating anomalous $WW\gamma$ couplings in $e^+e^-$ collisions},
journal = {Physics Letters B},
volume = {457},
number = {1},
pages = {193-201},
year = {1999},
issn = {0370-2693},
doi = {https://doi.org/10.1016/S0370-2693(99)00527-4},
url = {https://www.sciencedirect.com/science/article/pii/S0370269399005274},
author = {Debajyoti Choudhury and Jan Kalinowski and Anna Kulesza},
}

@article{Tizchang:2020tqs,
    author = "Tizchang, Seddigheh and Etesami, Seyed Mohsen",
    title = "{Pinning down the gauge boson couplings in WW{\ensuremath{\gamma}} production using forward proton tagging}",
    eprint = "2004.12203",
    archivePrefix = "arXiv",
    primaryClass = "hep-ph",
    doi = "10.1007/JHEP07(2020)191",
    journal = "JHEP",
    volume = "07",
    pages = "191",
    year = "2020"
}

@misc{rizzo1999polarizationasymmetriesgammae,
      title={Polarization Asymmetries in $\gamma e$ Collisions and Triple Gauge Boson Couplings Revisited}, 
      author={Thomas G. Rizzo},
      year={1999},
      eprint={hep-ph/9907395},
      archivePrefix={arXiv},
      primaryClass={hep-ph},
      url={https://arxiv.org/abs/hep-ph/9907395}, 
}

@article{Yang:2012vv,
    author = "Yang, Daneng and Mao, Yajun and Li, Qiang and Liu, Shuai and Xu, Zijun and Ye, Ke",
    title = "{Probing W+W-gamma Production and Anomalous Quartic Gauge Boson Couplings at the CERN LHC}",
    eprint = "1211.1641",
    archivePrefix = "arXiv",
    primaryClass = "hep-ph",
    doi = "10.1007/JHEP04(2013)108",
    journal = "JHEP",
    volume = "04",
    pages = "108",
    year = "2013"
}

@article{Bobeth:2015zqa,
    author = "Bobeth, Christoph and Haisch, Ulrich",
    title = "{Anomalous triple gauge couplings from $B$-meson and kaon observables}",
    eprint = "1503.04829",
    archivePrefix = "arXiv",
    primaryClass = "hep-ph",
    reportNumber = "FLAVOUR(267104)-ERC-93, CERN-PH-TH-2015-041",
    doi = "10.1007/JHEP09(2015)018",
    journal = "JHEP",
    volume = "09",
    pages = "018",
    year = "2015"
}

@misc{Biswal:2014oaa,
    author = "Biswal, Sudhansu S. and Patra, Monalisa and Raychaudhuri, Sreerup",
    title = "{Anomalous Triple Gauge Vertices at the Large Hadron-Electron Collider}",
    eprint = "1405.6056",
    archivePrefix = "arXiv",
    primaryClass = "hep-ph",
    reportNumber = "TIFR-TH-14-15",
    month = "5",
    year = "2014"
}

@article{Baur:1989gh,
    author = "Baur, U. and Zeppenfeld, D.",
    title = "{Measuring the $W W \gamma$ Vertex in Single $W$ Production at $e p$ Colliders}",
    reportNumber = "CERN-TH-5362/89, MAD/PH/486",
    doi = "10.1016/0550-3213(89)90457-4",
    journal = "Nucl. Phys. B",
    volume = "325",
    pages = "253--274",
    year = "1989"
}

@article{Kumar:2015lna,
    author = "Kumar, Satendra and Poulose, Poulose",
    title = "{Probing $WW\gamma$ coupling through $e^- \gamma \rightarrow \nu_e W^-$ at ILC}",
    eprint = "1501.01380",
    archivePrefix = "arXiv",
    primaryClass = "hep-ph",
    doi = "10.1142/S0217751X15502152",
    journal = "Int. J. Mod. Phys. A",
    volume = "30",
    number = "36",
    pages = "1550215",
    year = "2015"
}

@article{Li:2017kfk,
    author = "Li, Ruibo and Shen, Xiao-Min and Wang, Kai and Xu, Tao and Zhang, Liangliang and Zhu, Guohuai",
    title = "{Probing anomalous $WW\gamma$ triple gauge bosons coupling at the LHeC}",
    eprint = "1711.05607",
    archivePrefix = "arXiv",
    primaryClass = "hep-ph",
    doi = "10.1103/PhysRevD.97.075043",
    journal = "Phys. Rev. D",
    volume = "97",
    number = "7",
    pages = "075043",
    year = "2018"
}

@article{Baur:1993ix,
    author = "Baur, U. and Errede, S. and Ohnemus, J.",
    title = "{Ratios of $W^\pm \gamma$ and $Z \gamma$ cross-sections: New tools in probing the weak boson sector at the tevatron}",
    eprint = "hep-ph/9304239",
    archivePrefix = "arXiv",
    reportNumber = "FSU-HEP-930322, UIUC-HEP-93-01, DTP-93-14",
    doi = "10.1103/PhysRevD.48.4103",
    journal = "Phys. Rev. D",
    volume = "48",
    pages = "4103--4120",
    year = "1993"
}

@article{Dawson:2013owa,
    author = "Dawson, Sally and Gupta, Sudhir Kumar and Valencia, German",
    title = "{CP violating anomalous couplings in $W\gamma$ and $Z\gamma$ production at the LHC}",
    eprint = "1304.3514",
    archivePrefix = "arXiv",
    primaryClass = "hep-ph",
    doi = "10.1103/PhysRevD.88.035008",
    journal = "Phys. Rev. D",
    volume = "88",
    number = "3",
    pages = "035008",
    year = "2013"
}

@article{Dawson:1996ge,
    author = "Dawson, S. and He, Xiao-Gang and Valencia, G.",
    title = "{CP violation in W gamma and Z gamma production}",
    eprint = "hep-ph/9609523",
    archivePrefix = "arXiv",
    reportNumber = "BNL-HET-SD-96-6, ISU-HET-96-2",
    doi = "10.1016/S0370-2693(96)01364-0",
    journal = "Phys. Lett. B",
    volume = "390",
    pages = "431--436",
    year = "1997"
}

@article{Gutierrez-Rodriguez:2019hek,
    author = {Guti{\'e}rrez-Rodr{\'\i}guez, A. and K{\"o}ksal, M. and Billur, A. A. and Hern{\'a}ndez-Ru{\'\i}z, M. A.},
    title = "{Probing model-independent limits on $W^+W^-\gamma$ triple gauge boson vertex at the LHeC and the FCC-he}",
    eprint = "1910.02307",
    archivePrefix = "arXiv",
    primaryClass = "hep-ph",
    doi = "10.1088/1361-6471/ab7ff9",
    journal = "J. Phys. G",
    volume = "47",
    number = "5",
    pages = "055005",
    year = "2020"
}

@article{Ari:2015tca,
    author = {Ar{\i}, V. and Billur, A. A. and {\.I}nan, S. C. and K{\"o}ksal, M.},
    title = "{Anomalous WW{\ensuremath{\gamma}} couplings with beam polarization at the Compact Linear Collider}",
    eprint = "1506.08998",
    archivePrefix = "arXiv",
    primaryClass = "hep-ph",
    doi = "10.1016/j.nuclphysb.2016.02.029",
    journal = "Nucl. Phys. B",
    volume = "906",
    pages = "211--230",
    year = "2016"
}

@article{Koksal:2019oqt,
    author = {K{\"o}ksal, M. and Billur, A. A. and Guti{\'e}rrez-Rodr{\'\i}guez, A. and Hern{\'a}ndez-Ru{\'\i}z, M. A.},
    title = "{Bounds on the non-standard $W^+W^-\gamma$ couplings at the LHeC and the FCC-he}",
    eprint = "1910.06747",
    archivePrefix = "arXiv",
    primaryClass = "hep-ph",
    doi = "10.1016/j.physletb.2020.135661",
    journal = "Phys. Lett. B",
    volume = "808",
    pages = "135661",
    year = "2020"
}

@article{Eboli:2024wgy,
    author = "{\'E}boli, O. J. P. and Ghosh, Tathagata and Martines, Matheus and Shil, Sujay",
    title = "{Triple gauge coupling analysis using boosted W{\textquoteright}s and Z{\textquoteright}s}",
    eprint = "2410.13072",
    archivePrefix = "arXiv",
    primaryClass = "hep-ph",
    doi = "10.1140/epjc/s10052-025-14801-x",
    journal = "Eur. Phys. J. C",
    volume = "85",
    number = "10",
    pages = "1094",
    year = "2025"
}

@article{Subba:2022czw,
    author = "Subba, Amir and Singh, Ritesh K.",
    title = "{Role of polarizations and spin-spin correlations of W's in e-e+{\textrightarrow}W-W+ at s=250{\,}{\,}GeV to probe anomalous W-W+Z/{\ensuremath{\gamma}} couplings}",
    eprint = "2212.12973",
    archivePrefix = "arXiv",
    primaryClass = "hep-ph",
    doi = "10.1103/PhysRevD.107.073004",
    journal = "Phys. Rev. D",
    volume = "107",
    number = "7",
    pages = "073004",
    year = "2023"
}

@article{Rahaman:2023pte,
    author = "Rahaman, Rafiqul and Subba, Amir",
    title = "{Probing top quark anomalous moments in W boson associated single top quark production at the LHC using polarization and spin correlation}",
    eprint = "2306.06889",
    archivePrefix = "arXiv",
    primaryClass = "hep-ph",
    doi = "10.1103/PhysRevD.108.055027",
    journal = "Phys. Rev. D",
    volume = "108",
    number = "5",
    pages = "055027",
    year = "2023"
}

@article{Subba:2023jia,
    author = "Subba, Amir and Singh, Ritesh K.",
    title = "{Sensitivity of polarizations and spin correlations of Z boson to anomalous neutral triple gauge couplings at lepton collider with polarized beams}",
    eprint = "2311.12170",
    archivePrefix = "arXiv",
    primaryClass = "hep-ph",
    doi = "10.1103/PhysRevD.109.055047",
    journal = "Phys. Rev. D",
    volume = "109",
    number = "5",
    pages = "055047",
    year = "2024"
}

@misc{Subba:2024ojy,
    author = "Subba, Amir and Singh, Ritesh K.",
    title = "{Anomalous triple gauge couplings in $e^-e^+\to 4j$: Role of polarizations, spin correlations and interference}",
    eprint = "2407.01322",
    archivePrefix = "arXiv",
    primaryClass = "hep-ph",
    month = "7",
    year = "2024"
}

@article{Subba:2025pos,
    author = "Subba, Amir and Singh, Ritesh K.",
    title = "{Bounds on SMEFT affecting multigauge and Higgs-gauge couplings using two and three body spin correlations in e-e+{\textrightarrow}2je-e+{\ensuremath{\mu}}E process}",
    eprint = "2509.01452",
    archivePrefix = "arXiv",
    primaryClass = "hep-ph",
    doi = "10.1103/ntgz-czmt",
    journal = "Phys. Rev. D",
    volume = "113",
    number = "3",
    pages = "035026",
    year = "2026"
}

@article{Subba:2025hkq,
    author = "Subba, Amir and Bhattacharya, Subhaditya and Sarkar, Abhik",
    title = "{Probing CP-violating Higgs-gauge couplings with Higgsstrahlung at e-e+ colliders}",
    eprint = "2511.18541",
    archivePrefix = "arXiv",
    primaryClass = "hep-ph",
    doi = "10.1103/973l-lj5c",
    journal = "Phys. Rev. D",
    volume = "113",
    number = "5",
    pages = "055018",
    year = "2026"
}

@misc{Subba:2025hxn,
    author = "Subba, Amir and Singh, Ritesh K.",
    title = "{Probing anomalous $W^-W^+\gamma/Z$ couplings in the SMEFT with $e^-e^+ \to W^-W^+ \to 4j/\ell^-\ell^+\slashed{E}$ channel}",
    eprint = "2512.06671",
    archivePrefix = "arXiv",
    primaryClass = "hep-ph",
    month = "12",
    year = "2025"
}

@misc{Eboli:2026mbq,
    author = "{\'E}boli, Oscar J. P. and Rahaman, Rafiqul and Subba, Amir",
    title = "{Probing anomalous quartic gauge couplings in same-sign $W$ boson scattering with polarization and spin correlation}",
    eprint = "2606.06436",
    archivePrefix = "arXiv",
    primaryClass = "hep-ph",
    month = "6",
    year = "2026"
}

@article{Rahaman:2016pqj,
    author = "Rahaman, Rafiqul and Singh, Ritesh K.",
    title = "{On polarization parameters of spin-1 particles and anomalous couplings in $e^+e^-\rightarrow ZZ/Z\gamma $}",
    eprint = "1604.06677",
    archivePrefix = "arXiv",
    primaryClass = "hep-ph",
    doi = "10.1140/epjc/s10052-016-4374-4",
    journal = "Eur. Phys. J. C",
    volume = "76",
    number = "10",
    pages = "539",
    year = "2016"
}

@article{Rahaman:2017qql,
    author = "Rahaman, Rafiqul and Singh, Ritesh K.",
    title = "{On the choice of beam polarization in $e^+e^-\rightarrow ZZ/Z\gamma $ and anomalous triple gauge-boson couplings}",
    eprint = "1703.06437",
    archivePrefix = "arXiv",
    primaryClass = "hep-ph",
    doi = "10.1140/epjc/s10052-017-5093-1",
    journal = "Eur. Phys. J. C",
    volume = "77",
    number = "8",
    pages = "521",
    year = "2017"
}

@article{Rahaman:2019mnz,
    author = "Rahaman, Rafiqul and Singh, Ritesh K.",
    title = "{Probing the anomalous triple gauge boson couplings in $e^+e^-\to W^+W^-$ using $W$ polarizations with polarized beams}",
    eprint = "1909.05496",
    archivePrefix = "arXiv",
    primaryClass = "hep-ph",
    doi = "10.1103/PhysRevD.101.075044",
    journal = "Phys. Rev. D",
    volume = "101",
    number = "7",
    pages = "075044",
    year = "2020"
}

@article{Rahaman:2019lab,
    author = "Rahaman, Rafiqul and Singh, Ritesh K.",
    title = "{Unravelling the anomalous gauge boson couplings in $ZW^\pm$ production at the LHC and the role of spin-$1$ polarizations}",
    eprint = "1911.03111",
    archivePrefix = "arXiv",
    primaryClass = "hep-ph",
    doi = "10.1007/JHEP04(2020)075",
    journal = "JHEP",
    volume = {04},
    pages = {075},
    year = {2020}
}

@misc{Booth:1993af,
    author = "Booth, Michael J.",
    title = "{The Electric dipole moment of the W and electron in the Standard Model}",
    eprint = "hep-ph/9301293",
    archivePrefix = "arXiv",
    reportNumber = "EFI-93-01",
    month = "1",
    year = "1993"
}

@article{Shabalin:1978rs,
    author = "Shabalin, E. P.",
    title = "{Electric Dipole Moment of Quark in a Gauge Theory with Left-Handed Currents}",
    reportNumber = "ITEP-31-1978",
    journal = "Sov. J. Nucl. Phys.",
    volume = "28",
    pages = "75",
    year = "1978"
}

@article{Pospelov:1991zt,
    author = "Pospelov, M. E. and Khriplovich, I. B.",
    title = "{Electric dipole moment of the W boson and the electron in the Kobayashi-Maskawa model}",
    journal = "Sov. J. Nucl. Phys.",
    volume = "53",
    pages = "638--640",
    year = "1991"
}

@article{Khriplovich:1993js,
    author = "Khriplovich, I. B. and Pospelov, M. E.",
    title = "{Magnetic quadrupole moment of W boson in Kobayashi-Maskawa model}",
    eprint = "hep-ph/9310342",
    archivePrefix = "arXiv",
    reportNumber = "BUDKER-INP-1993-88, BINP-93-88",
    doi = "10.1016/0550-3213(94)90075-2",
    journal = "Nucl. Phys. B",
    volume = "420",
    pages = "505--516",
    year = "1994"
}

@article{Montano:2005kt,
    author = "Montano, J. and Ramirez-Zavaleta, F. and Tavares-Velasco, G. and Toscano, J. J.",
    title = "{Electric dipole and magnetic quadrupole moments of the W boson via a CP-violating HWW vertex in effective Lagrangians}",
    eprint = "hep-ph/0510255",
    archivePrefix = "arXiv",
    doi = "10.1103/PhysRevD.72.115009",
    journal = "Phys. Rev. D",
    volume = "72",
    pages = "115009",
    year = "2005"
}

@article{Salzman1966,
  author = {Salzman, F. and Salzman, G.},
  title = {Electric dipole moments of fermi particles},
  journal = {Il Nuovo Cimento A (1965-1970)},
  year = {1966},
  volume = {41},
  number = {4},
  pages = {443--455},
  month = {2},
  issn = {1826-9869},
  doi = {10.1007/BF02754814},
  url = {https://doi.org/10.1007/BF02754814}
}

@article{Jahedi:2024wnw,
    author = "Jahedi, Sahabub and Lahiri, Jayita and Subba, Amir",
    title = "{Optimal sensitivity of anomalous charged triple gauge couplings through W boson helicity at the e$^{+}$e$^{−}$ colliders}",
    eprint = "2411.13664",
    archivePrefix = "arXiv",
    primaryClass = "hep-ph",
    doi = "10.1007/JHEP07(2025)280",
    journal = "JHEP",
    volume = "07",
    pages = "280",
    year = "2025"
}

@article{Panico:2018hal,
    author = "Panico, Giuliano and Pomarol, Alex and Riembau, Marc",
    title = "{EFT approach to the electron Electric Dipole Moment at the two-loop level}",
    eprint = "1810.09413",
    archivePrefix = "arXiv",
    primaryClass = "hep-ph",
    reportNumber = "DESY-18-185",
    doi = "10.1007/JHEP04(2019)090",
    journal = "JHEP",
    volume = "04",
    pages = "090",
    year = "2019"
}

@misc{CMS:2026lik,
    author = "Gevorgyan, Arzunik and others",
    collaboration = "CMS",
    title = "{Search for anomalous couplings in WW and WZ production with single-lepton final states in proton-proton collisions at $\sqrt{s}$ = 13 TeV}",
    eprint = "2608.18245",
    archivePrefix = "arXiv",
    primaryClass = "hep-ex",
    reportNumber = "CMS-SMP-25-001, CERN-EP-2026-210",
    month = "8",
    year = "2026"
}

@article{Boudjema:1990dv,
    author = "Boudjema, F. and Hagiwara, Kaoru and Hamzaoui, C. and Numata, K.",
    title = "{Anomalous moments of quarks and leptons from nonstandard W W gamma couplings}",
    reportNumber = "KEK-TH-254, UDEM-LPN-TH-33, UT-KOMABA-90-18, KEK-PREPRINT-90-39",
    doi = "10.1103/PhysRevD.43.2223",
    journal = "Phys. Rev. D",
    volume = "43",
    pages = "2223--2232",
    year = "1991"
}

@article{Novales-Sanchez:2007rsw,
    author = "Novales-Sanchez, H. and Toscano, J. J.",
    title = "{Effective Lagrangian approach to fermion electric dipole moments induced by a CP-violating WW gamma vertex}",
    eprint = "0712.2008",
    archivePrefix = "arXiv",
    primaryClass = "hep-ph",
    doi = "10.1103/PhysRevD.77.015011",
    journal = "Phys. Rev. D",
    volume = "77",
    pages = "015011",
    year = "2008"
}

@article{Roussy:2022cmp,
    author = "Roussy, Tanya S. and others",
    title = "{An improved bound on the electron{\textquoteright}s electric dipole moment}",
    eprint = "2212.11841",
    archivePrefix = "arXiv",
    primaryClass = "physics.atom-ph",
    doi = "10.1126/science.adg4084",
    journal = "Science",
    volume = "381",
    number = "6653",
    pages = "adg4084",
    year = "2023"
}
\end{document}